\documentclass[11pt,reqno]{article}
\usepackage{fullpage}

\pdfoutput=1

\usepackage[T1]{fontenc}
\usepackage[utf8]{inputenc}
\usepackage[english]{babel}

\usepackage{amsmath,amssymb,bm,graphicx,psfrag}
\usepackage{mathtools}
\usepackage{bbold}
\usepackage{amsthm}

\usepackage{mdframed}
\usepackage{booktabs}
\usepackage{longtable}
\usepackage[ruled,vlined]{algorithm2e}
\usepackage{xcolor}
\usepackage{soul}
\usepackage{authblk}
\usepackage{enumitem}
\usepackage{ragged2e}
\usepackage{todonotes}

\usepackage{tikz}
\usepackage{pgfplots}
\pgfplotsset{compat=1.16}
\usetikzlibrary{decorations.pathreplacing,angles,quotes}
\usetikzlibrary{matrix,arrows,decorations.pathmorphing}
\usetikzlibrary{decorations.markings}
\usetikzlibrary{quantikz}
\usepackage{tikz-cd}
\usepackage[all]{xypic}

\usepackage[unicode=true]{hyperref}

\theoremstyle{plain}
\newtheorem{Theorem}{Theorem}[section]
\newtheorem{Corollary}[Theorem]{Corollary}
\newtheorem{Lemma}[Theorem]{Lemma}
\newtheorem{Prop}[Theorem]{Proposition}

\theoremstyle{definition}
\newtheorem{Definition}[Theorem]{Definition}
\newtheorem{Ex}[Theorem]{Example}
\newtheorem{Rem}[Theorem]{Remark}

\def\on{\operatorname}

\newcommand{\eff}{\operatorname{eff}}

\newcommand{\LP}{\operatorname{LP}}

\newcommand{\tightoverset}[2]{%
\mathop{#2}\limits^{\vbox to -.5ex{\kern-0.25ex\hbox{$#1$}\vss}}}

\newcommand{\NS}{\operatorname{NS}}

\newcommand{\idy}{\text{id}}

\newcommand{\one}{\mathbb{1}}
\newcommand{\zero}{\mathbb{0}}

\newcommand{\loc}{{{\on{L}}}}

\newcommand{\Tr}{\operatorname{Tr}}

\newcommand{\Herm}{\operatorname{Herm}}

\newcommand{\maxw}{\circ}

\newcommand{\LC}{\operatorname{LC}}

\newcommand{\set}[1]{\ensuremath{ \lbrace #1 \rbrace }}

\newcommand{\Span}[1]{\ensuremath{ \langle #1 \rangle }}
\newcommand{\norm}[1]{\ensuremath{ \| #1 \| }}

\newcommand{\Corr}{\operatorname{Corr}}

\newcommand{\cone}{\text{cone}}
\newcommand{\conv}{\text{conv}}

\newcommand{\ZZ}{\mathbb{Z}}

\newcommand{\RR}{\mathbb{R}}
\newcommand{\CC}{\mathbb{C}}
\newcommand{\NN}{\mathbb{N}}

\newcommand{\PP}{\mathbb{P}}

\newcommand{\oO}{\mathcal{O}}

\newcommand{\hH}{\mathcal{H}}

\newcommand{\kK}{\mathcal{K}}

\definecolor{bluegray}{rgb}{0.4, 0.6, 0.8}
\definecolor{turquoise}{rgb}{0.2, 0.7, 0.6}
\definecolor{darkgreen}{rgb}{0.4, 0.6, 0.3}
\definecolor{orange}{rgb}{0.74,0.38,0.09}
\definecolor{darkpurple}{rgb}{0.6, 0.1, 0.6}

\usepackage{soul}  

\newcounter{commcount}
\usepackage{xcolor}
\usepackage{todonotes}

 \title{Classical simulation of universal measurement-based quantum computation using multipartite Bell scenarios}

\author[1]{Cihan Okay}

\author[1, 2]{Atak Talay Y\" ucel}

\author[1,3]{Selman Ipek}

\affil[1]{{\small{Department of Mathematics, Bilkent University, Ankara, Turkey}}}
\affil[2]{{\small{Department of Computer Engineering, Bilkent University, Ankara, Turkey}}}
\affil[3]{\small{{Institut f\"{u}r Theoretische Physik, Leibniz Universit\"{a}t Hannover, Hannover, Germany}}}

\begin{document}

\maketitle

\begin{abstract}
  We introduce a novel classical simulation algorithm based on non-signaling polytopes of multipartite Bell scenarios, capable of simulating universal measurement-based quantum computation with single-qubit Pauli measurements.
In our model, the simultaneous presence of
non-stabilizer states
and entanglement is necessary for quantum speedup. The region of quantum states that can be efficiently simulated includes the Bell polytope
and extends beyond the reach of current sampling algorithms based on phase-space methods.
\end{abstract}

\section{Introduction}

\noindent
Understanding the reasons behind quantum computational advantage is a fundamental problem in theoretical quantum computing. One direct approach is to determine which quantum computations can be classically simulated in an efficient way. A cornerstone result in this direction is the Gottesman--Knill theorem \cite{gottesman1998heisenberg,aaronson2004improved}, which states that \emph{stabilizer circuits} can be efficiently classically simulated.

{Quantum computation with magic states (QCM) extends stabilizer circuits to universal quantum computation by supplying magic states that are processed through adaptive Pauli measurements~\cite{bravyi2005universal}. This model has proven fruitful for studying the resources responsible for quantum advantage, with} phase-space simulation methods playing a distinguished role. In this framework, classical simulation proceeds by sampling from a quasi-probability representation, such as the discrete Wigner functions~\cite{galvao2005discrete,cormick2006classicality,veitch2012negative,delfosse2015wigner,raussendorf2017contextuality} or their generalizations~\cite{raussendorf2020phase,zurel2025simulation,zurel2024efficient}. Negativity in the representation is then viewed as a necessary resource for quantum computational speedup.

{From a foundational perspective}, quantum advantage arises from intrinsic features of quantum theory such as contextuality and entanglement.
 {For systems with odd local dimensions, the foundational and classical simulation perspectives are closely aligned: in the relevant stabilizer setting, contextuality and Wigner negativity are equivalent notions of nonclassicality}~\cite{howard2014contextuality, howard2017application,delfosse2017equivalence}.
 {For qubits, however, this alignment breaks down owing to state-independent contextuality, which is already present within the stabilizer framework, as illustrated by the Mermin square paradox~\cite{mermin1993hidden}.}
 {Thus, retaining all stabilizer operations as computational primitives prevents the same correspondence between noncontextuality, nonnegative representations, and efficient classical simulation that holds in odd local dimensions.}

{This obstruction motivates two approaches: generalizing the phase-space representation while retaining the computational primitives of QCM, or restricting those primitives. The first approach extends phase space beyond the Wigner framework~\cite{raussendorf2020phase,zurel2025simulation,zurel2024efficient}, providing a more uniform treatment of computations in even and odd local dimensions while realigning negativity with the cost of classical simulation.}
 A recent tableau algorithm developed in~\cite{ipek2025phase}, and implemented in~\cite{CNCSim}, based on such generalized phase-space methods \cite{raussendorf2020phase} extends the Gottesman--Knill tableau technique \cite{gottesman1998heisenberg,aaronson2004improved} for stabilizer circuits to broader circuit families.

A key breakthrough in this direction was the construction of a probabilistic representation of universal quantum computation on a finite state space consisting of the vertices of the \emph{Pauli polytopes}, also known as the $\Lambda$ polytopes~\cite{zurel2020hidden,zurel2024hidden}. These polytopes can be viewed as a natural extension of stabilizer theory through a {convex-geometric dual construction}. Owing to their probabilistic structure, negativity vanishes from the simulation picture, and identifying the quantum property responsible for computational hardness remains an open challenge.

{The second approach restores} the alignment between negativity and contextuality as computational resources for qubit {quantum computation} by imposing operational restrictions~\cite{delfosse2015wigner,bermejo2017contextuality} {that eliminate state-independent contextuality from the computational model.
 Recovering universal quantum computation then requires a more expressive resource state.}

In this work, we ask: ``Are there other polytopes that provide a probabilistic representation of universal quantum computation?''
{Our affirmative answer to this question comes from combining
the two approaches above: we restrict the computational primitives
to single-qubit Pauli measurements and construct a polyhedral
probabilistic representation adapted to these operations.
We introduce a new simulation polytope, which we call the
\emph{local Pauli polytope}.}
 This locality is reflected in the dual formulation, which uses local stabilizer states.
  The local Pauli polytope strictly contains the $\Lambda$ polytope and can be identified with the non-signaling polytope of a multipartite Bell scenario \cite{brunner2014bell}. This identification reconnects foundational notions---such as the Bell polytope and extremal contextual non-signaling distributions---with the study of quantum advantage through
   probabilistic sampling algorithms.

{The operational restrictions underlying this construction also distinguish our computational model from the QCM setting described above, with fundamental consequences for the complexity of classical simulation.}
Our approach adopts the measurement-based quantum computation (MBQC) model, where entanglement is pre-established in the resource state rather than generated during the computation. Specifically, the computation begins with an entangled magic state and proceeds via adaptive single-qubit Pauli measurements. This initial state can be aptly called a \emph{magic cluster state}, {as it resembles a standard cluster state with some of the qubits rotated into magic states.} A distinctive feature of this model is the use of \emph{destructive} measurements, where each measurement consumes the corresponding qubit. Our simulation framework is built around these local measurements and benefits from a dynamic structure: the destructive nature of the computation reduces the system size at each step, leading to simplified complexity in the simulation process.


 \begin{figure}[htbp]
    \centering
    \hspace{0.22cm}
    \includegraphics[width=0.45\textwidth]{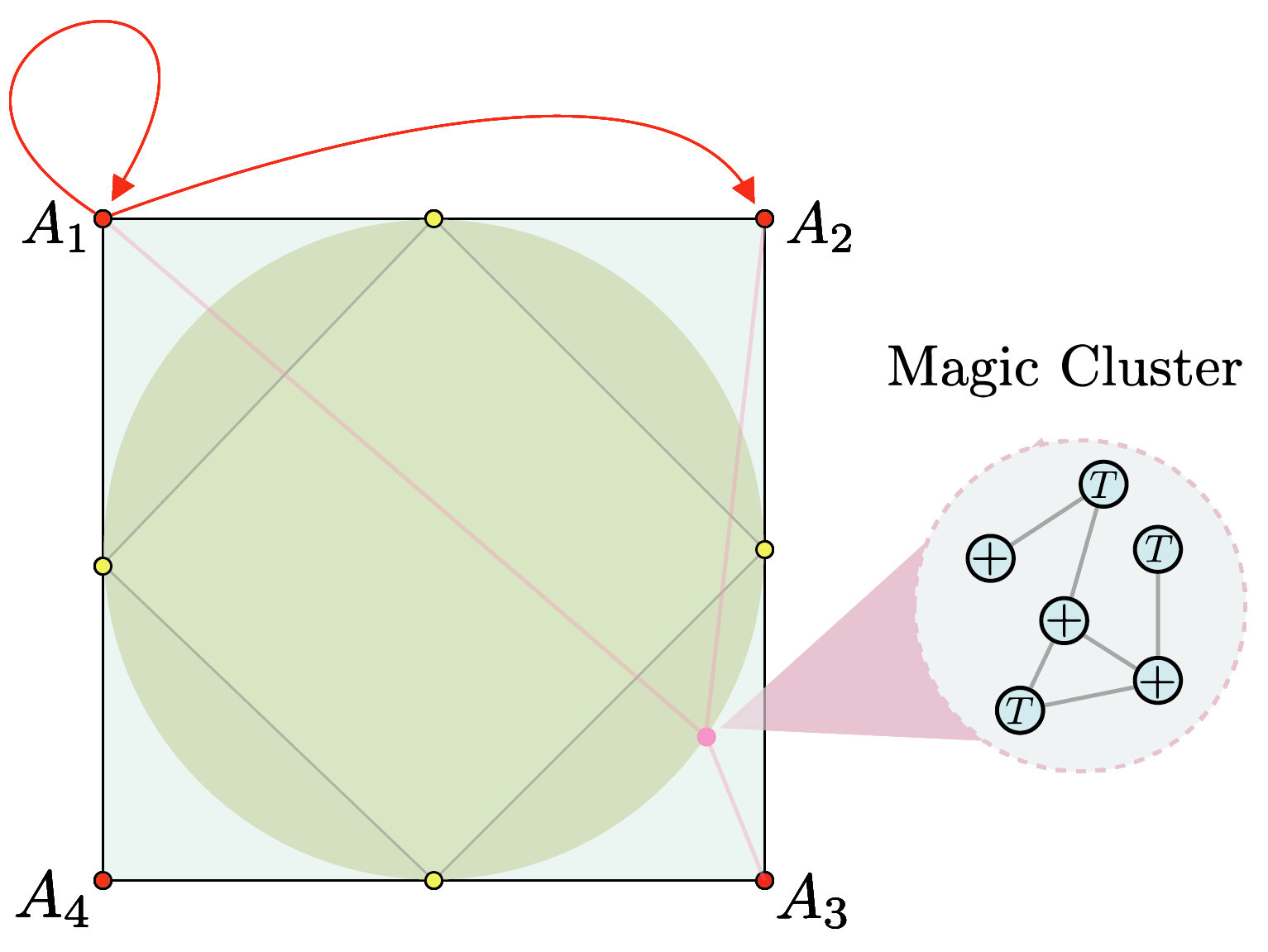}
    \caption{{Our computational model, MBPC,} utilizes both entanglement and magic as computational resources, operating through single-qubit Pauli measurements. Each such measurement sequence determines a local stabilizer state, represented by a yellow vertex. The corresponding dual polytope, shown with red vertices, serves as our simulation polytope. The resource state is expressed as a probabilistic mixture of these vertices and is updated probabilistically with each measurement.}
    \label{fig:fig}
\end{figure}

{The local Pauli simulator provides a systematic framework for identifying efficiently simulatable families of quantum states. New operator families with efficient representations and update rules can enlarge this sector, provided that the initial decomposition can also be sampled efficiently.}
By construction, the state space of our simulator strictly contains the $\Lambda$ state space, offering broader coverage {while exhibiting a simpler} geometric structure.

Our new local version refines the {geometric} investigation of {quantum advantage} with the following improvements:
\begin{enumerate}
    \item {The local Pauli polytope is defined by a proper subset of the
half-space inequalities defining $\Lambda$, reducing the number
of constraints in the vertex enumeration problem suggesting more tractable computational complexity.}
    \item {This} polytope is closed under tensor product, meaning that the tensor product of two vertices remains a vertex{---a crucial property absent in $\Lambda$ polytopes.}
    \item {The} computational model uses destructive measurements, where each step of the simulation reduces the number of qubits, thereby lowering computational cost{---in contrast to simulation methods based on QCM, such as the $\Lambda$ simulation.}
\end{enumerate}

{The computational model underlying the local Pauli simulator offers an example of an operationally restricted model in which contextuality and negativity align as resources for qubit computation, as demonstrated in~\cite{bermejo2017contextuality}.
 {Our framework connects this operational perspective with the geometry of non-signaling correlations, using both to guide the construction of classical simulation algorithms.}

\subsection{{Main results and organization}}
This paper introduces a new polyhedral classical simulator that extends the capabilities of existing simulators. A key novelty lies in shifting the computational model from {QCM} to measurement-based Pauli computation (MBPC), a variant of MBQC where entanglement is shifted to the initial resource state and computation proceeds via adaptive single-qubit Pauli measurements. The local nature of these measurements is reflected in the geometry of the underlying simulation polytope.
We adopt a unifying perspective on computational models involving destructive (e.g., MBQC) and non-destructive (e.g., QCM) measurements. This is formalized via a generalized framework of \emph{adaptive instruments}, which subsume both models. Our simulator is designed to simulate computations modeled by such instruments, thereby generalizing the $\Lambda$ simulation of \cite{zurel2020hidden}.

{Our central contribution is the development of \emph{polyhedral classical simulation}, which provides a probabilistic simulation framework for computations defined via adaptive instruments.}The key requirement for the simulation algorithm to {be correct} is a preservation property that characterizes how the instrument interacts with the underlying state polytopes (Definition \ref{def:preserved} in Section \ref{sec:model}).
This framework yields a family of simulation polytopes: for example, allowing all Pauli measurements recovers $\Lambda$, while restricting to local Pauli measurements yields the new \emph{local Pauli polytope} (Definition \ref{def:local Lambda}). Correctness of our simulation algorithm is proved in Theorem \ref{thm:preserves then CS correct}. We expect that variants can be built using various stabilizer and extended stabilizer theories \cite{van2013efficient,bermejo2014classical,webster2022xp}.

The geometry of our simulator offers advantages absent in prior constructions of quasiprobabilistic simulators.
\begin{enumerate}
\item {\bf Tensor decomposition} Notably, the local structure ensures that the set of vertices is closed under tensor product. We prove this using a generalized probabilistic theory (GPT) framework, where it follows from the behavior of extremal points under partial trace (Theorem \ref{thm:tensor} in Section \ref{sec:structural properties}). This tensor product property simplifies update rules and supports efficient simulation. $\Lambda$ lacks this feature.

\item {\bf Bell scenarios} On the foundational side, we identify the $n$-qubit local Pauli polytope with the non-signaling polytope of the $n$-partite Bell scenario, where each party has three binary-outcome measurements (Theorem \ref{thm:Lambda local is NS}). The familiar Bell polytope---corresponding to local or non-contextual correlations---marks a region of efficient classical simulation due to its tensor decomposition. In contrast to odd-dimensional systems where non-contextuality and efficient simulability align, our model admits contextual yet efficiently simulatable states. This follows from the inclusion of the $\Lambda$ polytope, and in particular, the CNC operators {\cite{raussendorf2020phase}}, within the local Pauli polytope. Any state within the CNC polytope but outside the Bell polytope exemplifies this phenomenon.

\item {\bf Improved efficiency} While simulators like those based on Wigner functions or more generally CNC operators can handle states outside their domain via quasi-probability distributions, the overhead is captured by the robustness measure, i.e., the minimal $1$-norm of such distributions. We show that the local Pauli polytope offers lower overheads compared to robustness measures from CNC-based simulations. This improvement is enabled by a new class of operators we introduce, termed \emph{locally closed operators}, which serve as the local analogs of CNC operators (Definition \ref{def:locally closed} in Section \ref{sec:locally closed operators}). We provide a detailed analysis of the resulting robustness measures (Table \ref{tab:comparison robustness} in {Section \ref{sec:comparison of efficiency})}.

\end{enumerate}

{
The comparison of the efficiency of the local Pauli simulator with that of the $\Lambda$ simulator depends on a comparison of their underlying computational models. We prove that any Clifford+$T$ circuit with $t$ $T$ gates can be efficiently converted into MBPC on a $t$-qubit resource state, reproducing the original output distribution up to efficient classical post-processing (Theorem~\ref{thm:t qubit resource} in Section~\ref{sec:reduced-resource}). This conversion result is crucial for comparing the simulators at the same resource-qubit count.
}

The remainder of the paper is organized as follows. In Section~\ref{sec:computational setting}, we introduce our computational model, {MBPC}, {and describe its basic properties.}
Section~\ref{sec:model} presents our polyhedral classical simulator for adaptive instruments and introduces our main construction, the \emph{local Pauli simulator}. We also prove the correctness of the simulation algorithm in this section.
In Section~\ref{sec:structural properties}, we demonstrate two important structural properties {using techniques from generalized probabilistic theories (GPT)}: that the tensor product of vertices remains a vertex, and that the local Pauli polytope coincides with the non-signaling polytope of a multipartite Bell scenario.
Section~\ref{sec:locally closed operators} introduces the notion of {locally closed operators}, along with two key subclasses: \emph{deterministic vertices}, which correspond to the vertices of the Bell polytope, and \emph{max-weight vertices}, which capture the computational power of nonlocal correlations.
{In Section~\ref{sec:locally closed vertex simulators}, we describe the update rules for locally closed operators and introduce the simulators corresponding to the deterministic and max-weight vertices.}
{Section~\ref{sec:local-robustness-efficiency}} compares the efficiency of our simulator to that of earlier polyhedral simulators by analyzing the transformations between different computational models.
We conclude in Section~\ref{sec:conclusion}.

\section{Computational setting}
\label{sec:computational setting}

\subsection{{Models of measurement-based computation}}
\label{sec:models of measurement-based computation}

{Our computational model combines the magic-state resources of the QCM model~\cite{bravyi2005universal} with the graph-state resources of MBQC~\cite{raussendorf2001one}. Both models rely on techniques from stabilizer theory~\cite{gottesman1998heisenberg,nielsen2001quantum}, which we now recall. Pauli operators are defined as tensor products of the single-qubit Pauli matrices $\one$, $X$, $Y$, and $Z$. The subgroup of unitaries generated by the Pauli operators is known as the \emph{Pauli group}, and the normalizer of the Pauli group, up to global phases, is called the \emph{Clifford group}. A quantum state is called a \emph{stabilizer state} if it can be written as $U\ket{0}^{\otimes n}$ for some Clifford unitary $U$, where $\ket{0}$ denotes the $+1$ eigenstate of $Z$.

A typical quantum circuit begins with the computational basis state $\ket{0}^{\otimes n}$, applies a unitary $U$, and terminates with measurements in the computational basis, i.e., in the Pauli $Z$-eigenbasis. Equivalently, by  using the Hadamard gate $H$, we can begin with the Pauli $X$-eigenstate $\ket{+}=H\ket{0}$ and terminate with Pauli $X$-measurements.
Throughout, we will work with \emph{approximately universal} quantum circuits admitting a Clifford$+T$ decomposition, i.e., circuits defined over the gate set $\{H_i,T_i,S_i,E_{ij}\}$. Among these gates, the Hadamard gate $H$, the phase gate $S$, and the controlled-$Z$ gate $E_{ij}$ generate the $n$-qubit Clifford group. By the celebrated Gottesman--Knill theorem~\cite{gottesman1998heisenberg,aaronson2004improved}, if $U$ is restricted to lie in the Clifford group, then the resulting circuit, called a \emph{stabilizer circuit}, is efficiently simulatable by probabilistic classical computation. To restore universality, the $T$ gates must be injected as resources. The resulting computational model is called the \emph{magic-state} model, or QCM~\cite{bravyi2005universal}, where these gates are injected using the magic state
$\ket{T}=T\ket{+}=(\ket{0}+e^{i\pi/4}\ket{1})/\sqrt{2}$.
The computation then proceeds by adaptive multi-qubit Pauli measurements interspersed with Clifford unitaries.

From the perspective of classical simulation, our goal is to sample from the output distribution determined by the Born rule probabilities. For this purpose, we are interested in quantum circuits that are \emph{computationally universal}, i.e., those that can reproduce the outcome probabilities of approximately universal quantum circuits~\cite{van2007fundamentals}. In QCM, the Clifford unitaries can be moved past the measurements, leaving a model that proceeds by measurements only~\cite{raussendorf2017contextuality}. Furthermore, the sequence of measurements can be made compatible while preserving computational universality; this variant is known as \emph{Pauli-based computation} (PBC)~\cite{bravyi2016trading}. Given an $n$-qubit quantum circuit with a unitary $U$ whose Clifford$+T$ decomposition contains $t$ $T$ gates, the corresponding PBC is initialized at $\ket{T}^{\otimes t}$ and proceeds by an adaptive sequence of compatible multi-qubit Pauli measurements~\cite{yoganathan2019quantum}.

The prototypical model of measurement-based computation is the \emph{one-way} model of~\cite{raussendorf2001one}. Its basic principle relies on a subclass of stabilizer states known as \emph{graph states}. Given a graph $G$ with vertex set $\{1,2,\dots,n\}$, the corresponding graph state has the form $E(G)\ket{+}^{\otimes n}$, where $E(G)$ is the product of the gates $E_{ij}$ over all edges $ij$ of the graph. Any stabilizer state can be put into graph-state form using the \emph{local Clifford group}, the subgroup generated by single-qubit Clifford unitaries~\cite{van2004graphical}. More precisely, given a stabilizer state $\ket{\psi}$, there exists a graph state and a local Clifford unitary $V$ such that $V\ket{\psi}=E(G)\ket{+}^{\otimes n}$. In measurement-based quantum computation (MBQC),
{a computation begins with a suitable graph state, such as a \emph{cluster state}}, and proceeds by a sequence of adaptive single-qubit measurements, typically in the $XY$ plane of the Bloch sphere.
}

\subsection{{Measurement-based Pauli computation}}
\label{sec:MBPC}

{In measurement calculus~\cite{danos2007measurement}, universal models of MBQC can be constructed from the fundamental pattern
\[
{\mathcal{J}}(\alpha) = X_2^{s_1} M_1(-\alpha) E_{12}.
\]
This pattern acts on the state {$\ket{\psi}\otimes \ket{+}$} by applying the entangling gate $E_{12}$, performs a measurement in the $XY$ plane with angle $-\alpha$ on the first qubit, and applies a Pauli $X$ correction to the second qubit {producing $J(\alpha)\ket{\psi}$}{, where $J(\alpha) = H \operatorname{diag}(1, e^{i\alpha})$.} Any quantum circuit expressed in the Clifford$+T$ gate set can be obtained by implementing $H$ using the ${\mathcal{J}}(0)$ pattern and the $Z$-rotation gate $R_Z(\alpha)$ {(up to a global phase)} using the composite pattern ${\mathcal{J}}(0){\mathcal{J}}(\alpha)$, with $\alpha=\pi/2$ and $\pi/4$ for the $S$ and $T$ gates, respectively.}

Our computational setting is {a variant of standard MBQC, which we call \emph{measurement-based Pauli computation} (MBPC).} In this model, a computation
\begin{itemize}
\item begins with a magic cluster state
\begin{equation}\label{eq:resource psi G U}
\ket{\psi_{G,U}} = E(G) T^{\otimes U} \ket{+}^{\otimes n}
\end{equation}
where $T^{\otimes U}$ acts by $T$ on a subset $U$ of qubits, and
\item proceeds with a sequence of adaptive single-qubit Pauli $X$ and $Y$ measurements.
\end{itemize}
{Together with the $X$ and $S$ corrections, t}his computational model
is {approximately} universal~\cite{danos2007pauli}.

{
  We are interested in weak simulation~\cite{vandenNest2010classical,pashayan2020estimation} of MBPC, namely sampling from the output distribution.
  For this purpose, the outcome-dependent $X$ and $S$ corrections can be absorbed into adaptive measurement choices and classical post-processing of the outcomes.
  The resulting model remains computationally universal, in the sense that it can reproduce the outcome probabilities of approximately universal quantum circuits~\cite{peres2023quantum}.
}

{We see that MBPC requires only the ability to perform single qubit Pauli $X$ and $Y$ measurements on a suitable resource state for quantum universality. As with other MBQC models requiring only local Pauli measurements, such as those based on so-called hypergraph states \cite{miller2016hierarchy,takeuchi2019quantum}, the cost of simpler measurements is typically an increase in the number of required qubits.  We return to this point in Section~\ref{sec:local-robustness-efficiency}, where we use techniques from the theory of graph states to reduce the qubit count of our resource states in order to facilitate a more direct comparison to PBC.}

\section{{Polyhedral classical simulation}}
\label{sec:model}

 \subsection{{General framework}}

Classical simulation algorithms based on quasi-probability distributions, e.g.,~\cite{veitch2012negative}, and their probabilistic versions~\cite{zurel2020hidden}, both rely on the same principle: sampling from a state space. This state space consists of a subset of an operator-theoretic polytope, typically its set of vertices. We will consider weak simulation of adaptive {quantum computations} using operator-theoretic polytopes, thereby generalizing earlier methods developed for simulating QCM~\cite{veitch2012negative,zurel2020hidden}.

For a uniform treatment of computational models based on destructive and non-destructive measurements, we employ the framework of \emph{instruments}~\cite{watrous2018theory}. {An instrument with outcome set $\Sigma$ is a collection of completely positive maps $\Phi^s\in \on{CP}(\hH,\kK)$, indexed by $s\in\Sigma$, such that $\sum_{s\in \Sigma} \Phi^s$ is a quantum channel.}
By a polytope, we mean a compact convex subset of real Euclidean space with finitely many extremal points, called vertices.
Let $P$ be a polytope in the space $\Herm_1(\hH)$ of trace-one Hermitian operators, and let $V(P)\subseteq P$ be a finite subset of operators that will serve as the state space associated with the polytope. The elements of this set will be denoted by $A_\alpha$, where $\alpha$ is the state label. We require that $P$ is the convex hull of this set.

\begin{Definition}\label{def:preserved}
Given polytopes $P\subseteq \Herm_1(\hH)$ and $Q\subseteq \Herm_1(\kK)$, we say that an instrument $\Phi$ \emph{preserves} the pair $(P,Q)$ if, for all $A\in P$ and $s\in \Sigma$, we have $\Tr(\Phi^s(A))\geq 0$ and the following two properties hold:
\begin{enumerate}
\item[(i)] If $\Tr(\Phi^s(A))>0$, then
\[
\frac{\Phi^s(A)}{\Tr(\Phi^s(A))}\in Q.
\]
\item[(ii)] If $\Tr(\Phi^s(A))=0$, then
\[
\Phi^s(A)=\zero.
\]
\end{enumerate}
\end{Definition}

If $\Phi$ preserves $(P,Q)$, then we can define an update map $q=\set{q_\alpha}_{\alpha\in V(P)}$ consisting of probability distributions
\begin{equation}\label{eq:update}
q_\alpha: V(Q)\times \Sigma \to \RR_{\geq 0}
\end{equation}
defined by Property {(i)}:
\[
\Phi^s({A_\alpha}) = \sum_{\beta\in V(Q)} q_\alpha(\beta,s) A_\beta
.\]
We say that the resulting classical simulation algorithm is {\it correct} if for every $A\in P$ expressed as
\[
A = \sum_{\alpha\in V(P)} p_A(\alpha) A_\alpha
\]
and for $s\in \Sigma$ such that $\Tr(\Phi^s(A))>0$ it holds that the post-measurement state  of the classical simulation coincides with {$\Phi^s(A)/\Tr(  \Phi^s(A) )$,} the quantum mechanical post-measurement state {when $A$ is a density operator}. {Equivalently, for all $s\in \Sigma$, the following equation holds:
\begin{equation}\label{eq:correctness}
\Phi^s(A) = \sum_\beta {\PP}({\beta,
s}) A_\beta
\end{equation}
where ${\PP}({\beta,
s})$ is the probability that outcome $s$ and state $\beta$ are obtained in the simulation.}

Our first main result is the generalization of operator-theoretic simulation results, such as those in~\cite{zurel2020hidden,zurel2024hidden}, to the setting of adaptive instruments. This yields a formulation of polyhedral classical simulation for quantum computations expressed within the adaptive-instrument model.

{An \emph{adaptive instrument} consists of instruments $\Phi_a$ labeled by inputs $a\in \Gamma$. Given another adaptive instrument $\Psi$, whose input set is the same as the outcome set of $\Phi$ and whose outcome set is $\Theta$, we can compose them to obtain the adaptive instrument $\Psi\star \Phi$ defined by
\[
(\Psi\star \Phi)_a^{sr} = \Psi_s^r \circ \Phi_a^s,
\]
with inputs $a\in\Gamma$ and outcomes $(s,r)$. A quantum computation can be specified by a sequence of adaptive instruments $\Phi_1,\cdots,\Phi_N$, where the input set of $\Phi_i$, for $2\leq i\leq N$, is the product of the outcome sets $\Sigma_1\times \cdots\times \Sigma_{i-1}$ of the previous instruments. Usually, the first instrument $\Phi_1$ is provided with a single input, as is typical in QCM. Such a sequence then implements an 
{adaptive instrument}
given by the adaptive composition
\[
\Phi=\bigl(\Phi_N\star(\Phi_{N-1}\star \cdots \star (\Phi_{2}\star \Phi_{1})\cdots)\bigr).
\]
One can think of this formalism as a generalized version of the measurement calculus for adaptive instruments.}

\begin{Theorem}\label{thm:preserves then CS correct}
The classical simulation algorithm in Algorithm~\ref{alg:Algorithm} is correct{, i.e., it reproduces the Born rule probabilities of the quantum computation.}
\end{Theorem}\begin{proof}
For an instrument $\Phi$ preserving a pair of polytopes $(P,Q)$, we check that Eq.~\eqref{eq:correctness} holds. We have
\[
\begin{aligned}
\Phi^s(A)
&= \sum_{\alpha\in V(P)} p_A(\alpha)\Phi^s(A_\alpha) \\
&= \sum_{\alpha:\, \Tr(\Phi^s(A_\alpha))>0}
p_A(\alpha)\sum_{\beta\in V(Q)} q_\alpha(\beta,s)A_\beta \\
&= \sum_{\beta\in V(Q)}
\sum_{\alpha:\, \Tr(\Phi^s(A_\alpha))>0}
q_\alpha(\beta,s)p_A(\alpha)A_\beta \\
&= \sum_{\beta\in V(Q)} {\PP}(\beta,s)A_\beta.
\end{aligned}
\]
Here Properties~(i) and~(ii) are used in the second equality. The probability ${\PP}(s)$ of obtaining outcome $s$ is obtained from the joint distribution ${\PP}(\beta,s)$ and agrees with the Born rule probability:
\[
\Tr(\Phi^s(A)) = \sum_{\beta\in V(Q)} {\PP}(\beta,s).
\]
For an adaptive instrument as in Algorithm~\ref{alg:Algorithm}, the classical simulation is correct at each step because the post-measurement states agree. {Note that the update map at step $i$ depends on the previous $i-1$ outcomes.} As a result, the algorithm produces probabilities ${\PP}(s_1,\ldots,s_N)$ that agree with those of the quantum-mechanical process.
\end{proof}

  \begin{algorithm}
\justifying
\begin{enumerate}[leftmargin=*]
\item Sample from the probability distribution $p_\rho$ to obtain the initial vertex label $\alpha_0\in V(P_0)$.
\item For $i=1,\ldots,N$, sample from the corresponding update map $q_{\alpha_{i-1}}^{(i)}$ to obtain a new vertex label $\alpha_i\in V(P_i)$ and an outcome $s_i\in \Sigma_i$. Output $s_i$ and update the vertex label $\alpha_{i-1}\mapsto \alpha_i$.
\end{enumerate}
\caption{\label{alg:Algorithm}
Classical algorithm based on the polytopes $P_0,\ldots,P_N$ for simulating a single run of a quantum computation consisting of a sequence $\Phi_1,\ldots,\Phi_N$ of adaptive instruments, where each $\Phi_i$ preserves the pair $(P_{i-1},P_i)$, initialized at $\rho\in P_0$.
}
\end{algorithm}

{The formalism of adaptive instruments and the corresponding polyhedral simulation can be depicted using diagrams that clearly separate the quantum and classical processes. Compositional properties of adaptive instruments and polyhedral simulators are detailed in \cite{okay2025polyhedral}.}

\subsection{Local Pauli simulator}
\label{sec:local pauli simulator}

{The first example of a polyhedral simulator whose state space contains all quantum states was introduced in~\cite{zurel2020hidden}. The $n$-qubit $\Lambda$ polytope is defined dually as the set of trace-one Hermitian operators with non-negative inner product with all stabilizer states. We refer to this polytope as the Pauli polytope, since operationally these inequalities amount to non-negativity under Pauli measurements. The resulting polyhedral simulator can reproduce the Born rule probabilities of a quantum computation expressed in the QCM model. In this section, we introduce the main object of interest of this paper, namely a local version of the Pauli polytope, and demonstrate that it serves as a simulation polytope for MBPC.}

A stabilizer state of the form $U\ket{0}^{\otimes n}$, where $U$ belongs to the subgroup of the Clifford group generated by local Clifford unitaries and permutations of the tensor factors, is called a \emph{local stabilizer state}.
{Equivalently, a local stabilizer state is a tensor product of single-qubit stabilizer states.}
There are exactly $2^n \times 3^n$ such $n$-qubit local stabilizer states. We denote the set of $n$-qubit local stabilizer states by $S_n^\loc$. {We sometimes prefer to represent a stabilizer state with the corresponding projector.}

\begin{Definition}\label{def:local Lambda}
The local {Pauli} polytope is defined by
$$
{\LP_n} =\set{A\in \Herm_1(\hH):\,  \Tr(\ket{\psi}\bra{\psi} A)\geq 0,\;\forall \ket{\psi} \in {S_n^\loc} }
$$
where $\hH=(\CC^2)^{\otimes n}$ is the $n$-qubit Hilbert space.
\end{Definition}

{Our second main result is the operational characterization of the local Pauli polytope.}
For the further analysis {of this polytope}, we need some preliminaries concerning the symplectic structure of stabilizer theory; see, e.g.,~\cite{zurel2024hidden}. Every Pauli operator can be written as
\[
T_a = i^{a_X\cdot a_Z} X^{(a_X)_1}Z^{(a_Z)_1}\otimes \cdots \otimes X^{(a_X)_n}Z^{(a_Z)_n},
\]
where $a\in E_n=\ZZ_2^n \times \ZZ_2^n$ and $a_X\cdot a_Z$ denotes the dot product of the bit strings, viewed modulo $4$ in the exponent. The projector onto the $(-1)^s$-eigenspace of $T_a$ is defined by
\[
\Pi_a^s = \frac{\one + (-1)^sT_a}{2}.
\]
To describe the commutation relations among Pauli operators, we define a symplectic form
\[
[\cdot,\cdot]:E_n\times E_n\to \ZZ_2
\]
by
\[
T_aT_b=(-1)^{[a,b]}T_bT_a.
\]
We say that $a$ and $b$ commute when $[a,b]=0$; otherwise, they anticommute.

A subspace $I\subseteq E_n$ is called isotropic if every pair of elements in $I$ commutes. A value assignment is a function $s:I \to \ZZ_2$ satisfying
\[
s(a+b)=s(a)+s(b)+\beta(a,b)
\]
for all $a,b\in I$, where $\beta$ is obtained from the product $T_aT_b=(-1)^{\beta(a,b)}T_{a+b}$. Given a subspace $J\subseteq I$, we write $s|_J$ for the restriction of $s$ to $J$. A stabilizer projector is defined by
\[
\Pi_I^s = \Pi_{a_1}^{s_1} \cdots \Pi_{a_k}^{s_k},
\]
where $I=\Span{a_1,\ldots,a_k}$ is an isotropic subspace with basis $a_1,\ldots,a_k$, and $s:I\to \ZZ_2$ is defined by $s_i=s(a_i)$. If $I$ is maximal, then the corresponding projector specifies a stabilizer state.

Next, we introduce the {symplectic theory for the} local version of stabilizer theory. We are primarily concerned with single-qubit Pauli operators $X_i,Y_i,Z_i$ acting on the $i$-th qubit. Let $x_i$ and $z_i$ denote the bit strings whose $i$-th and $(i+n)$-th entries, respectively, are $1$, with all other entries equal to $0$. Also let $y_i=x_i+z_i$. With this convention, we have $T_{x_i}=X_i$, $T_{y_i}=Y_i$, and $T_{z_i}=Z_i$. We call an element $a\in E_n$ \emph{local} if $a\in \set{x_i,y_i,z_i}$ for some $i$. A Pauli operator $T_a$ is called \emph{local} if $a$ is local. The corresponding measurement is referred to as a \emph{local} Pauli measurement.

Any element $a\in E_n$ has a unique local decomposition
\[
a=\sum_{i=1}^n \alpha_i a_i,
\]
where $a_i\in \set{x_i,y_i,z_i}$ if $\alpha_i\in \ZZ_2$ is nonzero, and $a_i=0$ otherwise. We say that $a$ and $b$ \emph{locally commute} if $[a_i,b_i]=0$ for all $1\leq i\leq n$, where $a_i$ and $b_i$ are the $i$-th terms in the local decompositions of $a$ and $b$, respectively. A subspace $I\subseteq E_n$ is called \emph{local isotropic} if every $a,b\in I$ locally commute. Note that $I$ is maximal local isotropic if and only if it is spanned by one nonzero local element on each qubit. A stabilizer projector $\Pi_I^s$ is called \emph{local} if $I$ is local isotropic, and a stabilizer state projector is called \emph{local} if $I$ is maximal local isotropic. Local stabilizer states correspond to maximal local isotropic subspaces.

\begin{Theorem}\label{thm:characterization Lambda local}
The local Pauli polytope $\LP_n$ is the largest polytope
$P\subseteq\Herm_1(\hH)$ such that the pair $(P,P)$ is preserved by
all non-destructive single-qubit Pauli measurements.
\end{Theorem}
\begin{proof}
{For a local Pauli operator $T_b$, let us write $\Phi_b^r (A) = \Pi_b^r A \Pi_b^r$ for the instrument  corresponding to the non-destructive measurement of this Pauli operator.}
{We first show that $(\LP_n,\LP_n)$ is preserved by all non-destructive
local Pauli measurements.}
{Take $A \in {\LP_n}$}.
We first observe that $\Tr(\Pi_{b}^{r}A)\geq 0$ since this trace is the marginalization of $\Tr(\Pi^s_IA)\geq 0$ for some local $I$ containing $b$ over value assignments satisfying $s(b)=r$. Property (i) follows from the probabilistic update rule of stabilizer states, {see, e.g., \cite{zurel2024hidden}}.
In detail,
assume that ${\Tr(\Phi_b^r(A))=} \Tr(\Pi_b^r A)>0$. We want to show that $\Tr(\Pi_{I}^{s}{\Phi_b^r(A)})\geq 0$ for all local stabilizer states.
Since $I$ is a maximal local isotropic, we have
\[
  \Pi_b^r \Pi_I^s \Pi_b^r = \begin{cases}
    \delta_{s(b),r} \Pi_I^s & \text{if } b\in I,\\
    \frac{1}{2}\Pi_{I(b)}^{s\ast r} & \text{if } b\notin I,
  \end{cases}
\]
where $I(b)=\Span{b}+\Span{b}^\perp\cap I$ {and
$s\ast r:I(b)\to\mathbb{Z}_2$ is the unique value assignment that
restricts to $s$ on $\Span{b}^\perp\cap I$ and satisfies $(s\ast r)(b)=r$.}
Thus,
\begin{align}
\Tr(\Pi_I^s{\Phi_b^r(A)})
&= \Tr(\Pi_b^r\Pi_I^s\Pi_b^r A) \nonumber\\
&=
\begin{cases}
\delta_{s(b),r}\Tr(\Pi_I^s A), & \text{if } b\in I,\\[1ex]
\dfrac{1}{2}\Tr(\Pi_{I(b)}^{s\ast r}A), & \text{if } b\notin I.
\end{cases}
\label{eq:trace Arb}
\end{align}
Therefore this expression is either zero or is (up to a positive constant) equal to {$\Tr(\Pi_I^s A)$ or} $\Tr(\Pi_{I(b)}^{s\ast r}A)$.
{Since both $\Pi_I^s$ and $\Pi_{I(b)}^{s\ast r}$ are local stabilizer states and $A \in \LP_n$, both traces are nonnegative.
Hence, ${\Phi_b^r(A)}$ belongs to ${\LP_n}$ after normalization.}

For Property (ii), assume $\Tr(\Pi_b^r A)=0$.
{From \eqref{eq:trace Arb}, we have $\Tr(\Pi_I^s{\Phi_b^r(A)})\geq 0$ for all local stabilizer states.
Moreover, since $\sum_s \Pi_I^s = \one$, we have
\[
  \sum_{s} \Tr(\Pi_I^s{\Phi_b^r(A)}) = \Tr\left( \left(\sum_{s} \Pi_I^s\right) {\Phi_b^r(A)} \right) = \Tr({\Phi_b^r(A)}) = \Tr(\Pi_b^r A) = 0.
\]
Hence, $\Tr(\Pi_I^s{\Phi_b^r(A)})=0$ for all local stabilizer states. Since these states span $\Herm(\hH)$, we conclude that ${\Phi_b^r(A)}=\zero$.
}

For the converse, let $P\subseteq\Herm_1(\hH)$ be any polytope such that
$(P,P)$ is preserved by all non-destructive local Pauli
measurements, and let $A\in P$.
For any local maximal isotropic  $I = \Span{a_1, a_2, \cdots, a_n} \subseteq E_n$ and any
value assignment $s: I \to \ZZ_2$ we can write
$\Pi_I^{s}$ as
a
product
$\Pi_I^{s} = \Pi_{a_1}^{s_1}\Pi_{a_2}^{s_2}\cdots\Pi_{a_n}^{s_n}$ where $s_i=s(a_i)$, {and each $a_i$ is local}. Therefore,
\[
\Tr(\Pi_I^{s}A) = \Tr(\Pi_{a_1}^{s_1}\cdots\Pi_{a_n}^{s_n}A\Pi_{a_n}^{s_n}\cdots\Pi_{a_1}^{s_1}).
\]
Now there are two possibilities: either $\Tr(\Pi_{a_n}^{s_n}A\Pi_{a_n}^{s_n}) >0$ or it is zero.
If it is zero, then $\Tr(\Pi_I^{s}A) = 0$ {by Property (ii)}.
Otherwise, {define the normalized post-measurement operator
\[A_{a_n}^{s_n} = \Pi_{a_n}^{s_n}A\Pi_{a_n}^{s_n} / \Tr(\Pi_{a_n}^{s_n}A).\]
By Property (i), $A_{a_n}^{s_n}$ {belongs to} $P$}{. W}e can repeat this {process} at most $n$ times to get the desired result.
If any of the intermediate traces are zero, then we have $\Tr(\Pi_I^{s}A) = 0$. Otherwise, $\Tr(\Pi_I^{s}A)$ is equal to the product of $n$ positive numbers. {Therefore} $\Tr(\Pi_I^{s}A) \ge 0$, and  $A\in {\LP_n}$.
\end{proof}

\begin{Corollary}\label{cor:destructive}
    The local Pauli polytope $\LP_n$ is the largest polytope $P \subseteq \Herm_1(\hH)$ such that the pair $(P,\LP_{n-1})$ is preserved by all destructive single-qubit Pauli measurements.
  \end{Corollary}
  \begin{proof}
    We first show that any preserved polytope $P \subseteq \Herm_1(\hH)$ must be contained in $\LP_n$.
    Let $A \in P$ and let $\Pi_I^s$ be an arbitrary $n$-qubit local stabilizer state. By definition, $\Pi_I^s$ factors into a tensor product of single-qubit stabilizer states. Without loss of generality, assume the destructive measurement acts on the $n$-th qubit, and write $\Pi_I^s = \Pi_{I'}^{s'} \otimes \Pi_{a_n}^{s_n}$, where $\Pi_{I'}^{s'}$ is a local stabilizer state on the first $n-1$ qubits.
    Let $\bar{\Phi}_{a_n}^{s_n} = \Tr_n\circ \Phi_{a_n}^{s_n}$ denote the destructive measurement instrument. We have the identity:
    \begin{equation*}
      \Tr(\Pi_I^s A) = \Tr\bigl(\Pi_{I'}^{s'}\,\Tr_n(\Pi_{a_n}^{s_n}A\,\Pi_{a_n}^{s_n})\bigr) = \Tr\bigl(\Pi_{I'}^{s'}\,\bar{\Phi}_{a_n}^{s_n}(A)\bigr).
    \end{equation*}
    Let $p = \Tr(\bar{\Phi}_{a_n}^{s_n}(A))${. S}ince $P$ is preserved, we have $p\geq 0$.
    If $p = 0$, Property~(ii) of Definition~\ref{def:preserved} implies $\bar{\Phi}_{a_n}^{s_n}(A) = \zero$, yielding $\Tr(\Pi_I^s A) = 0$.
    If $p > 0$, Property~(i) guarantees that $\bar{\Phi}_{a_n}^{s_n}(A)/p \in \LP_{n-1}$. Since $\Pi_{I'}^{s'}$ is a local stabilizer state on $n-1$ qubits,
    its inner product with an element of $\LP_{n-1}$ is non-negative. Therefore, $\Tr(\Pi_I^s A) \geq 0$.
    Thus, $A$ has non-negative overlap with all local stabilizer states, and we conclude that $A \in \LP_n$.

    Conversely, notice that the partial trace over the $i$-th qubit induces an affine map of polytopes $\LP_n\to \LP_{n-1}$. For simplicity, take $i=n$. For a local $(n-1)$-qubit stabilizer state $\Pi_I^s$, we have
    \[
    \langle \Pi_I^s,\Tr_n(A) \rangle
    =
    \langle \Pi_I^s\otimes \one,A \rangle
    \geq 0,
    \]
    where $\langle\cdot,\cdot\rangle$ denotes the Hilbert--Schmidt inner product. Indeed, we can write $\one=\Pi_a^0+\Pi_a^1$ for any $a\in\{x_n,y_n,z_n\}$, so $\Pi_I^s\otimes \one$ is a sum of local stabilizer projectors.

    Let $\bar\Phi_b^r=\Tr_i\circ \Phi_b^r$ denote the destructive measurement instrument associated with the local Pauli measurement $b$ on qubit $i$. Then, for $A\in \LP_n$, we have
    \[
    \Tr(\bar\Phi_b^r(A))=\Tr(\Phi_b^r(A))\geq 0.
    \]
    Properties~(i) and~(ii) for $\Phi_b^r$ hold by Theorem~\ref{thm:characterization Lambda local}. Applying the partial trace $\Tr_i$ gives the corresponding properties for $\bar\Phi_b^r$, since $\Tr_i$ maps $\LP_n$ into $\LP_{n-1}$.
  \end{proof}

{An important consequence of {Theorem~\ref{thm:characterization Lambda local} and its destructive version Corollary \ref{cor:destructive}} is that Algorithm~\ref{alg:Algorithm} provides a classical simulation of universal quantum computation in the MBPC model. In the non-destructive case, the simulation polytopes are chosen as $P_i=\LP_n$ for all $i=0,\ldots,N$. On the other hand, Corollary~\ref{cor:destructive} gives a simulation algorithm for the destructive case, where the polytopes are chosen as $P_i=\LP_{n-i}$, assuming that one qubit is measured destructively at each step.}

A similar characterization holds for the $\Lambda$ polytopes, which are defined analogously to the local Pauli polytopes, except that all stabilizer states are used in Definition~\ref{def:local Lambda}. {More precisely, let $S_n$ denote the set of $n$-qubit stabilizer states. Then $\Lambda_n$ consists of trace-one Hermitian operators $A$ satisfying $\Tr(\ket{\psi}\bra{\psi}A)\geq 0$ for all $\ket{\psi}\in S_n$. It follows immediately from the definitions that $\Lambda_n \subseteq \LP_n$.}
{Similar techniques can be used} to show that the $\Lambda$ polytope is the largest polytope in the space of trace-one Hermitian operators that is preserved under all Pauli measurements. This invariance property is the key ingredient underlying the simulation algorithm in~\cite{zurel2020hidden}. In this sense, the local Pauli polytope can be viewed as a local version of the $\Lambda$ polytope---a geometric manifestation of the locality of the measurements in the computational model being simulated.

\section{{Structural properties of the local Pauli polytope}}
\label{sec:structural properties}


We investigate two key structural features of the local Pauli polytope that are central to understanding its role in classical simulation. First, we examine its behavior under the tensor product operation, where we demonstrate a favorable compatibility property distinguishing it from the $\Lambda$ polytope. Second, we establish a correspondence between the local Pauli polytope and the non-signaling polytope of a multipartite Bell scenario.
We analyze these features within the framework of generalized probabilistic theories (GPT)~\cite{barrett2007information}, which provides a natural setting for studying the geometric and operational properties of our model.

\subsection{Background on GPTs}

{We recall some basic notions of GPTs following the presentation of \cite{plavala2023general}.} A GPT is specified by a state space $K$ {given} by a  convex, closed, and bounded {nonempty} subset of {the} real Euclidean space $V$, which we take to be finite-dimensional.
 To a state space $K$
 {one can associate a linear effect algebra $E(K)$ consisting of
{affine functions} $f:K\to\RR$ satisfying $0\leq f(x)\leq 1$ for all $x\in K$ with {an order unit} given by the function $1_K\in E(K)$ constant at $1$.}
 Conversely,
 it is possible to recover a state space $S(E)$ from a linear effect algebra $E$.
 The state space $S(E(K))$ can be identified with $K$, up to affine isomorphism.

In the GPT setting effects and states can be used to obtain the empirical probabilities of a given experimental setup according to
\[
p(e\lvert x) := e(x),
\]
where $e\in E(K)$ is an effect and $x\in K$ is a state. By construction these probabilities are non-negative. In quantum theory the state space is precisely the set of density operators acting on a finite-dimensional Hilbert space and the effects are elements of a positive operator-valued measure (POVM) \cite{watrous2018theory}. Indeed, we refer to a measurement in a GPT as any finite collection of effects
\[
\{e_a\}_{a}\subseteq E(K),
\quad \text{such that} \quad
\sum_{a} e_a = 1_K.
\]
The corresponding empirical probabilities clearly form a probability distribution.


\subsection{Multipartite GPTs}


Within the GPT framework it is possible to construct valid \textit{bipartite} state spaces $K_{1}\tilde \otimes K_{2}$ from subsystem state spaces $K_{1}$ and $K_{2}$ \cite{plavala2023general}.
 {In the GPT context, the tensor product is not uniquely defined, but there {is} instead a range of possible choices with two edge cases:}
\begin{itemize}
\item The minimal tensor product:
$$
K_1 \dot\otimes K_2 = \conv(\set{x\otimes y:\,x\in K_1,\,y\in K_2}).
$$

\item The maximal tensor product:
$$
K_1 \hat\otimes K_2 = S( E(K_1) \dot\otimes E(K_2) )
$$
where $E(K_1) \dot\otimes E(K_2)$ is the minimal tensor product of effect algebras.
\end{itemize}
Given state spaces $K_{1}$ and $K_{2}$ we can then also define the partial trace:
$$
{\Tr_1 : K_{1}\tilde \otimes K_{2} \to K_{2}}
$$
where $\Tr_1= 1_{K_1}\otimes \idy$. Partial trace on the other factor is defined similarly.

In a general bipartite GPT the decomposition of the unit $1_{K}$ into measurement effects is not unique. This guarantees that the empirical probabilities using separable effects satisfy the principle of nonsignaling \cite{barrett2007information}.

{\Lemma{
\label{lem:gpt-nonsignaling}
Let $x\in K_1\tilde\otimes K_2$ be a bipartite GPT state. Let
$\{e_a\}_a,\{\tilde e_{\alpha}\}_{\alpha}\subseteq E(K_1)$ be two measurements
on $K_1$, and let $\{f_b\}_b\subseteq E(K_2)$ be a measurement on $K_2$. Define
\[
p(a,b):=(e_a\otimes f_b)(x),
\qquad
\tilde p(\alpha,b):=(\tilde e_{\alpha}\otimes f_b)(x).
\]
Then
\[
\sum_a p(a,b)
=
f_b(\Tr_1(x))
=
\sum_{\alpha}\tilde p(\alpha,b).
\]
}}

\noindent
{\em{{Proof.}}}
The proof is straightforward \cite{janotta2014generalized}.
\hfill$\square$\par
\medskip

For multiple subsystems, the maximal and minimal tensor products are associative \cite{plavala2023general}. The $n$-fold minimal tensor product of a state space $K$ is therefore given by
\[
K^{\dot \otimes n} := \conv\left (\{x_{1}\otimes \cdots \otimes x_{n}~:~x_{i}\in K\}\right ),
\]
while the $n$-fold {maximal} tensor product is
\[
K^{\hat \otimes n} := S\left (E(K)\dot \otimes \cdots \dot \otimes E(K)\right ),
\]
which follows since
\[
E(K)\dot \otimes \cdots \dot \otimes E(K) = E(K\hat \otimes \cdots \hat \otimes K).
\]
The corresponding empirical probabilities using effects from $E(K\hat \otimes \cdots \hat \otimes K)$ in this more general setting still satisfy the principle of nonsignaling.

\subsection{Tensor structure of LP polytopes}

Since the states used in our classical simulation correspond to vertices of $\LP$, the enumeration of these extremal states becomes a central problem. Our main result in this section is a partial characterization of $\LP$ vertices using their intrinsic tensor structure. The result holds for arbitrary bipartite GPT state spaces.

{Recall that for a state space $K\subseteq V$ we say that $x\in K$ is an extreme point of $K$ if for every $y,z\in K$ and $\lambda \in (0,1)$ then $x = \lambda y +(1-\lambda) z$ implies that $x = y = z$. Denoting the set of extreme states of $K$ by $\text{ext}(K)$, we then have that $K = \conv(\text{ext}(K))$ \cite{plavala2023general}.}

{\Prop{Let $y\in K_{1}$ and $z\in K_{2}$ be extreme points of $K_{1}$ and $K_{2}$, respectively, and let $K_{1}\tilde \otimes K_{2}$ be a bipartite state space. Then the composite state $y \otimes z$ is an extreme point of $K_{1}\tilde \otimes K_{2}$.\label{pro:tensor-vertices-gpt}
}
}

\vspace{1em}
\noindent
{\em{{Proof.}} }
Let $x = y \otimes z$ and note that, by definition, $x\in K_{1}\tilde \otimes K_{2}$. Thus we can write $x$ as a convex combination
\[x = \sum_{\alpha}p(\alpha)v_{\alpha},\]
where $v_{\alpha}$ are extreme points of $K_{1}\tilde \otimes K_{2}$. Taking the partial trace on the first side of the partition gives $\Tr_{1}(x) = z$ which by the linearity of the partial traces yields
\[z = \Tr_{1}(x) = \sum_{\alpha}p(\alpha)\Tr_{1}(v_{\alpha}).\]
Since $z$ is an extreme point of $K_{2}$ this implies that $\Tr_{1}(v_{\alpha}) = z$ for all $\alpha$ for which $p(\alpha) > 0$. We will now use a fundamental result from \cite{plavala2023general} about GPTs: if some $u\in K_{1}\tilde \otimes K_{2}$ is such that $\Tr_{2}(u)=v$ is an extreme point of $K_{1}$ then $u=v \otimes w$ for some $w\in K_{2}$. This implies that $v_{\alpha} = w_{\alpha} \otimes z$ for some $w_{\alpha}\in K_{1}$. Taking the partial trace on the other factor we obtain $y = \Tr_{2}(x) = \sum_{\alpha}p(\alpha)w_{\alpha}$. From this we then deduce that $w_{\alpha} = y$. Thus for each $\alpha$ such that $p(\alpha) > 0$ we have $v_{\alpha} = y\otimes z$, which proves the result.
\hfill$\square$\par
\medskip

{
\begin{Lemma}\label{lem:positive-cone-LP}
The cone of positive affine functions on the local Pauli polytope can be identified with the cone of rank $1$ local stabilizer projectors:
\[
A(\LP_n)^+
=
\on{cone}(S_n^\loc).
\]
\end{Lemma}
\begin{proof}
The containment $\on{cone}(S_n^\loc)\subseteq A(\LP_n)^+$ is immediate by the definition of the state space $\LP_n$. The converse inclusion follows from Farkas' lemma in the form of \cite[Corollary 4.3]{bertsimas1997introduction} and the fact that the cone of $\LP_n$ can be identified with
\[
C_n = \{ A \in \Herm(\hH) \colon \Tr(\Pi A)\geq 0,\; \Pi\in S_n^\loc \}.
\]
The containment $\cone(\LP_n)\subseteq C_n$ is clear. For the converse, take $A\in C_n$. We have
\[
\Tr(A) = \Tr(\one A) = \sum_{j} \Tr(\Pi_jA)\geq 0
\]
where $\{\Pi_j\}_j$ are the rank $1$ projectors associated with a local stabilizer orthonormal basis of $\hH$, giving a resolution of the identity. Thus, if $\Tr(A)=0$, then each $\Tr(\Pi_jA)=0$. For every fixed $\Pi\in S_n^\loc$, choose a local stabilizer orthonormal basis containing $\Pi$ and apply the same argument to conclude that $\Tr(\Pi A)=0$. Since the local stabilizer projectors span $\Herm(\hH)$, it follows that $A=0$, and hence $A\in \cone(\LP_n)$. If $\Tr(A)>0$, then $\bar A:= A/\Tr(A)$ is of trace $1$ and thus belongs to $\LP_n$. Therefore, $A$ is in the cone.

Now, given $A\in A(\LP_n)^+$, we have $\Tr(AB)\geq 0$ for all $B\in C_n$. Thus, we apply Farkas' lemma to $C_n$ to conclude that any Hermitian operator $A$ satisfying $\Tr(AB)\geq 0$ for all $B\in C_n$ belongs to the cone of $S_n^\loc$.
\end{proof}
}

\begin{Theorem}\label{thm:tensor}
Given vertices $A_{\alpha_i}\in {\LP_{n_i}}$, where $i=1,2$, the tensor product $A_{\alpha_1}\otimes A_{\alpha_2}$ is a vertex of ${\LP_{n_1+n_2}}$.
\end{Theorem}
\begin{proof}
{The proof follows immediately from Proposition~\ref{pro:tensor-vertices-gpt} once we show that the $(n_1+n_2)$-qubit local Pauli polytope is given by the maximal tensor product of the $n_1$-qubit and $n_2$-qubit local Pauli polytopes:
\begin{equation}\label{eq:max tensor}
{\LP_{n_1+n_2} = \LP_{n_1} \hat\otimes  \LP_{n_2}}.
\end{equation}
Using Lemma~\ref{lem:positive-cone-LP}, we observe that the cone of positive affine functions factorizes as a minimal tensor product
\[
\begin{aligned}
A(\LP_{n_1+n_2})^+
&=
\on{cone}(S_{n_1+n_2}^\loc)\\
&=
\on{cone}
\set{
\Pi_1\otimes\Pi_2:
\Pi_i\in S_{n_i}^\loc
}\\
&=
A(\LP_{n_1})^+\dot\otimes A(\LP_{n_2})^+.
\end{aligned}
\]
The maximal tensor product of state spaces can be expressed as the dual of this minimal tensor product. Again using Lemma~\ref{lem:positive-cone-LP}, we obtain
$$
\begin{aligned}
\LP_{n_1}\hat\otimes\LP_{n_2}
&=
\set{
A\in\Herm_1(\hH):
\Tr(A(B_1\otimes B_2))\geq0
\text{ for all }
B_i\in A(\LP_{n_i})^+
}\\
&=
\set{
A\in\Herm_1(\hH):
\Tr(A\Pi)\geq0
\text{ for all }
\Pi\in S_{n_1+n_2}^\loc
}\\
&=
\LP_{n_1+n_2}.
\end{aligned}
$$}
\end{proof}

\subsection{Bell scenarios}

Next, we identify the local Pauli polytope with the nonsignaling polytope of a multipartite Bell scenario~\cite{brunner2014bell}. Let $\NS_{n}$ denote the non-signaling polytope of the $(n,3,2)$ Bell scenario, i.e., $n$-parties, $3$ measurements per party, and  binary outcomes {for each measurement}.

Our identification uses the fact that $\LP_{n}$ is the $n$-fold max tensor product
\[
\LP_{n} = \LP_{1} \hat \otimes \cdots \hat \otimes \LP_{1},
\]
which follows from {Eq.~\eqref{eq:max tensor} and}
the associativity of the maximal tensor product \cite{plavala2023general}. The $i$-th party can perform the single qubit Pauli measurements labeled by $x_i,y_i,z_i$ regarded as elements of $E_{n}$ and yielding the outcome $s_{i}$. For any GPT state $A\in \LP_{n}$ the empirical probabilities are then given by
\begin{eqnarray}\label{eq:nonsignaling-gpt}
p_{a_{1}\cdots a_{n}}^{s_{1}\cdots s_{n}} := \Tr(A\Pi_{a_{1}\cdots a_{n}}^{s_{1}\cdots s_{n}})
\quad
\text{where}
\quad
\Pi_{a_{1}\cdots a_{n}}^{s_{1}\cdots s_{n}} = {\prod_{i=1}^{n}\Pi_{a_{i}}^{s_{i}}}
\end{eqnarray}
is an extremal effect in $E(\LP_{n})$ and corresponds to a local stabilizer state.

\begin{Theorem}\label{thm:Lambda local is NS}
The {local Pauli} polytope {$\LP_{n}$} can be identified with the non-signaling polytope $\NS_n$ of the $(n,3,2)$ Bell scenario.
\end{Theorem}

\begin{proof}
Let $\phi:\LP_n\to \NS_n$ be the map sending $A\in \LP_n$ to the family of
empirical probabilities
\[
p_{a_1\cdots a_n}^{s_1\cdots s_n}
=
\Tr\left(A\Pi_{a_1\cdots a_n}^{s_1\cdots s_n}\right),
\]
as in Eq.~(\ref{eq:nonsignaling-gpt}).
{The map \(\phi\) is affine and injective because the local stabilizer projectors span the space of Hermitian operators.}
Lemma~\ref{lem:gpt-nonsignaling} implies that this family is nonsignaling since $\LP_n$ is the $n$-fold maximal tensor product of {$\LP_1$'s}. Thus $\phi(\LP_n)\subseteq \NS_n$. Conversely, let $p\in \NS_n$ and for each $a\in E_n$ denote by $e_{a}$ the corresponding expectation value with respect to $p$. Here we set $e_0=1$. Nonsignaling guarantees that the marginals are consistent and thus that the expectation values are well-defined. Introduce now
\[
A_{p}
=
\frac{1}{2^n}
\left(
\one+\sum_{a\in E_n{\setminus}\set{0}} e_a T_a
\right).
\]
Owing to the moment formula \cite{klyshko1996bell} we have
\[
\Tr\left(A_{p}\Pi_{a_1\cdots a_n}^{s_1\cdots s_n}\right)
=
p_{a_1\cdots a_n}^{s_1\cdots s_n}.
\]
Thus $A_{p}$ assigns nonnegative probabilities to every local stabilizer effect and hence $A_{p}\in \LP_{n}$ for all $p\in \NS_{n}$.
This implies that {$\phi$ is surjective}.
{Hence \(\phi\) is an affine isomorphism.}
\end{proof}

\section{Locally closed operators}
\label{sec:locally closed operators}

There is an important class of well-understood vertices of the {Pauli} polytopes called the closed non-contextual (CNC) vertices~\cite{raussendorf2020phase,zurel2020hidden}.
{{These operators admit efficient representations and update rules, enabling efficient simulation of quantum states whose convex decompositions into CNC operators can be sampled efficiently.} They extend the canonical Gottesman--Knill tableau algorithm to a phase-space tableau algorithm~\cite{ipek2025phase}. In this section,} we introduce their local counterparts {as suitable candidates for describing the efficiently simulatable sector of the local Pauli simulator}.

{We begin by recalling the corresponding definitions in the CNC case from~\cite{raussendorf2020phase}. A subset $\Omega\subseteq E_n$ is called \emph{closed} if, whenever $a,b\in \Omega$ and $[a,b]=0$, we have $a+b\in \Omega$. It is called \emph{non-contextual} if it admits a \emph{value assignment}, that is, a function $\gamma:\Omega\to \ZZ_2$ satisfying $\gamma(a+b)=\gamma(a)+\gamma(b)+\beta(a,b)$ whenever $[a,b]=0$. A subset $\Omega$ is called \emph{closed non-contextual} (CNC) if it is both closed and non-contextual.}

\begin{Definition}\label{def:locally closed}
A {nonempty} subset $\Omega\subseteq E_n$ is called {\it locally closed} if
$a,b\in \Omega$  locally {commute} then $a+b\in \Omega$.
A function $\gamma:\Omega\to \ZZ_2$ is called a {\it local value assignment}, if $\gamma(a+b)=\gamma(a)+\gamma(b)$ for all $a,b\in \Omega$ that locally commute.
We call, {$(\Omega,\gamma)$ consisting of} a locally closed set
together with {a} {local value assignment},
a {\it local pair}.
\end{Definition}

Given a local pair $(\Omega,\gamma)$, the associated \emph{locally closed operator} is defined by
\begin{equation}\label{eq:LC operator}
A_\Omega^\gamma = \frac{1}{2^n} \sum_{a\in \Omega} (-1)^{\gamma(a)} T_a.
\end{equation}
It is straightforward to verify that this operator belongs to $\LP_n$. {The class of l}ocally closed operators 
{contains }
the class of CNC operators \cite{raussendorf2020phase}.

\subsection{{Deterministic vertices}}
\label{sec:deterministic-vertices}

 The first class of locally closed vertices that we identify consists of the well-known \emph{deterministic vertices} of the non-signaling polytope $\NS_n${, which we observed is affinely isomorphic to $\LP_n$; see Theorem~\ref{thm:Lambda local is NS}}.
In the single-qubit case, which is equivalent to the eight-state model~\cite{wallman2012non}, these vertices are the locally closed operators of deterministic type given by
\[
A^{rst} = \frac{\one + (-1)^r X + (-1)^s Y+(-1)^t Z}{2}.
\]
For $n$ qubits, a deterministic vertex is specified by a local value assignment $\gamma:E_n\to \ZZ_2$ and has the tensor decomposition
\begin{equation}\label{eq:deterministic}
A_{E_n}^\gamma = A^{r_1s_1t_1} \otimes A^{r_2s_2t_2} \otimes \cdots \otimes A^{r_ns_nt_n},
\end{equation}
where $(r_i,s_i,t_i)=(\gamma(x_i),\gamma(y_i),\gamma(z_i))$. {We write $\on{DP}_n$ for the \emph{deterministic polytope}, defined as the convex hull of the deterministic vertices.}
When $\gamma$ is linear, these constructions coincide with the {qubit version of the} phase-point operators used in {the construction of Gross'} discrete Wigner function~\cite{gross2006hudson,raussendorf2017contextuality}.

{Indeed, for qubits it is known that the Wigner function is not covariant under the full Clifford group. Point operators are defined by $A_0=\frac{1}{2^n}\sum_{v\in E_n} T_v$ and, for $v\in E_n$, by $A_v=T_vA_0T_v$. In terms of deterministic vertex operators, in the single-qubit case, writing $v=(x,z)$, we have $A_v = A^{z(x\oplus z)x}$. For multiple qubits, $A_v$ is the tensor product of these single-qubit operators once $v$ is expressed in the standard binary representation. The Hadamard gate acts by $H A^{z(x\oplus z)x} H = A^{x(x\oplus z\oplus 1)z}$, moving a point operator of parity $0$ outside this family to its conjugate of parity $1$.

Let us write $p(\gamma) {= (p_1, \ldots, p_n)}\in \ZZ_2^n$ for the parity of a deterministic vertex operator, {where $p_i = \gamma(x_i) + \gamma(y_i) + \gamma(z_i) \pmod 2$ is the parity of the $i$-th single-qubit factor}. Operators of parity $p$ can be obtained from the standard point operators by the formula
\[
A_v(p) = \left( \bigotimes_{i=1}^n H^{p_i} \right) A_v \left( \bigotimes_{i=1}^n H^{p_i} \right).
\]
Each set $\{ A_v(p) : v\in E_n \}$ constitutes an orthogonal basis of the space of Hermitian operators. The set of deterministic vertex operators is precisely the union of these disjoint families of bases.} 

Finally, we remark that the stabilizer polytope is not contained in the deterministic polytope for $n\geq 3$. For $n=1$, all quantum states are contained in the deterministic polytope, since it coincides with the whole local Pauli polytope, namely the $8$-state model. For $n=2$, the fact that all stabilizer states are contained in $\on{DP}_n$ was proved in~\cite{ipek2023degenerate}. {For $n\geq 3$, the $n$-qubit GHZ state provides an example of a stabilizer state that lies outside the deterministic polytope~\cite{Mermin1990Unified,Mermin1990Extreme}.}




\subsection{{Max-weight vertices}}
\label{sec:power of correlations}

We now turn to a second, structurally rich class of vertices, which arise from the computational power of correlations rooted in non-locality across all partitions of the parties. These vertices generalize the computational advantage exemplified by the well-known Mermin star configuration involving the 3-qubit GHZ state~\cite{anders2009computational}. A key insight from~\cite{hoban2011generalized} is that such vertices of non-signaling polytopes in Bell scenarios can be effectively analyzed through their associated correlation polytopes. In what follows, we investigate this class in detail and identify a subclass of extremal points---called \emph{max-weight vertices}---that exhibit non-classical correlations not captured by bipartite linear functions.

Non-signaling distributions on an $n$-partite Bell scenario with $c$ measurements and $d$ possible outcomes for each party, {where $c$ and $d$ are prime,} can be studied by considering the resulting correlations \cite{hoban2011generalized}.
The polytope {$\Corr_n$} of correlations consists of stochastic maps from the set $\ZZ_c^n$ of measurements to the set $\ZZ_d$ of outcomes.
More explicitly, a point in this polytope is given by  a tuple $p=(p_{a_1\cdots a_n}^s)_{a_i\in \ZZ_c,s\in \ZZ_d}$ of real numbers satisfying non-negativity $p_{a_1\cdots a_n}^s\geq 0$ and  normalization $\sum_{s}p_{a_1\cdots a_n}^s=1$ conditions.
It turns out that the vertices of this polytope are given by the functions $f:\ZZ_c^n\to \ZZ_d$.
There is
an affine
map of polytopes
$$
c:\NS_n \to {\Corr_n}
$$
obtained by computing the correlations associated to a non-signaling distribution. This map is
{crucial}
in analyzing the vertices of the non-signaling polytope.

There are two classes of vertices detected by this map. A function $f$ is called {\it $n$-partite linear} if it can be written as $f(s)=\sum_{i=1}^n g_i(s_i)$ for some functions $g_i:\ZZ_c\to \ZZ_d$.
{The images of deterministic vertices under $c$ are precisely the vertices of $\Corr_n$ corresponding to $n$-partite linear functions.}
A function $f$ is called {\it bipartite linear} if there exists a partition of $\set{1,2,\cdots,n}$ into two disjoint subsets $A$ and $B$ such that $f(s)=g(s_A)+h(s_B)$ for some functions $g$ and $h$ of the form $\ZZ_c^{|A|}\to \ZZ_d$ and  $\ZZ_c^{|B|}\to \ZZ_d$, respectively. The inputs $s_A$ and $s_B$ are the tuples obtained from $s$ by restricting to the partitions. Then the fundamental result of \cite{hoban2011generalized} is the following:
{The non-signaling distribution $p\in \NS_n$ defined by}
\begin{equation}\label{eq:p Hoban}
p_{a_1\cdots a_n}^{s_1\cdots s_n}
=
\begin{cases}
d^{1-n} & \text{if } \sum_{i=1}^n s_i = f(a_1,\ldots,a_n), \\
0 & \text{otherwise}
\end{cases}
\end{equation}
is a vertex of $\NS_n$ if and only if $c(p)$ is the vertex of {$\Corr_n$} corresponding to a function $f$ that is not $n$-bipartite linear.

Now, we specialize to the case $c=3$ and $d=2$ to make a connection with locally closed operators.
{The {\it weight} $w(a)$ of an element in $a\in E_n$ is the number of non-zero elements in its local decomposition. We will write {$E^\maxw_n$} for the set of elements of weight $n$.
{The locally closed set we will consider is $\set{0}\sqcup E_n^{\maxw}$.}
By the identification in Theorem \ref{thm:Lambda local is NS},
non-signaling distributions in Eq.~(\ref{eq:p Hoban}) correspond to
locally closed operators
{specified by functions $\gamma:E_n^{\maxw}\to\ZZ_2$
that are not $n$-bipartite linear, under the identification}
\begin{equation}\label{eq:identificatoin to Z3n}
E_n^{\maxw}\cong\ZZ_3^n.
\end{equation}
{The function $\gamma$ specifies the max-weight correlations;
its domain does not contain the zero Pauli label.
Its extension $\bar\gamma$ to $\{0\}\sqcup E_n^{\maxw}$,
defined by $\bar\gamma(0)=0$, is a local value assignment.
We use the abbreviation
\[
A_{E_n^{\maxw}}^\gamma
:=
A_{\{0\}\sqcup E_n^{\maxw}}^{\bar\gamma}.
\]}

\begin{Lemma}\label{lem:maximal vs bipartite}
The set of maximal local pairs $({\set{0}\sqcup E_n^{\maxw},\bar \gamma})$ is in bijective correspondence with the set of functions $\gamma:\ZZ_3^n\to \ZZ_2$ that are not $n$-bipartite linear.
\end{Lemma}
\begin{proof}
{We define the local closure of a subset $S\subseteq E_n$ to be the
smallest locally closed set containing $S$.}
{For $a=\sum_{i=1}^n a_i$ in its local decomposition, define
$\on{supp}(a)=\{i:a_i\neq 0\}$.
The case $n=1$ is immediate, so assume $n\geq 2$.}
For a subset $A\subseteq \set{1,2,\cdots,n}$ we will write
$E_k^{(A)}$, where $k=|A|$, for the subspace in $E_n$ generated by
$\set{x_i,z_i:\,i\in A}$.
{Write $F_A=\{a\in E_k^{(A)}:\on{supp}(a)=A\}$.}
Given a partition $\set{1,2,\cdots,n}=A\cup B$, where both $A$ and $B$
are non-empty, we will write $\Omega(A,B)$ for the local closure of
{$F_A\cup F_B$}. Explicitly, we have
\[
\Omega(A,B)=\{0\}\sqcup F_A\sqcup F_B\sqcup E_n^{\maxw}.
\]
We claim that the smallest locally closed sets
{strictly containing $\{0\}\sqcup E_n^{\maxw}$}
are all of this form.
This can be seen by taking a nonzero element
$a\in E_n{\setminus}E_n^{\maxw}$ and considering the local closure of
$\set{a}\cup {E_n^{\maxw}}$.
{Set $A=\on{supp}(a)$ and let $B$ be its complement.
For every $b\in F_B$, the max-weight element $a+b$ locally commutes
with $a$, so the closure contains $b$. For any $u\in F_A$, the
max-weight element $u+b$ locally commutes with $b$, so the closure
also contains $u$.}
Thus the local closure coincides with $\Omega(A,B)$.
{Moreover, any element of $\Omega(A,B)$ outside
$\{0\}\sqcup E_n^{\maxw}$ has support $A$ or $B$ and generates the
same closure, establishing minimality.}

We finish by observing that $\gamma$ is $n$-bipartite linear if and
only if {$\bar\gamma$ extends to a local value assignment
$\widehat\gamma$ on some $\Omega(A,B)$}.
{Indeed, any extension $\widehat\gamma$ satisfies
$\gamma(u+v)=\widehat\gamma(u)+\widehat\gamma(v)$
for $u\in F_A$ and $v\in F_B$.
Conversely, a decomposition $\gamma(u+v)=g(u)+h(v)$ defines a local
value assignment $\widehat\gamma$ on $\Omega(A,B)$ by retaining
$\bar\gamma$ on $\{0\}\sqcup E_n^{\maxw}$ and assigning $g$ on $F_A$
and $h$ on $F_B$.}
\end{proof}

{In particular, under the bijection $E_n^\circ\cong\ZZ_3^n$, every function $\gamma:\ZZ_3^n\to\ZZ_2$ that is not $n$-bipartite linear defines a vertex of $\LP_n$ via Eq.~\eqref{eq:LC operator}.}

\subsection{Locally closed polytope}
\label{sec:locally closed polytope}

{We define the locally closed polytope {$\LC_n$} to be the convex hull of the locally closed operators.}
{It is an open question whether}
every vertex of this polytope is a vertex of the local {Pauli} polytope.
For $2$-qubits, the vertex enumeration problem for the corresponding non-signaling polytope $\NS_{2}$ is solved in \cite{jones2005interconversion}.
By further numerical verification one observes that
these vertices  are all locally closed operators and
fall into five orbits under the action of combinatorial automorphisms of the polytope. See Table~\ref{tab:ns232_orbits_verbose} for further details.

\begin{table}[ht]
\centering
\footnotesize
\setlength{\tabcolsep}{6.5pt}
\renewcommand{\arraystretch}{1.4}
\begin{tabular}{c c ccccccccccccccc}
\hline
Orbit & Size & $X_1$ & $Y_1$ & $Z_1$ & $X_2$ & $Y_2$ & $Z_2$ & $XX$ & $XY$ & $XZ$ & $YX$ & $YY$ & $YZ$ & $ZX$ & $ZY$ & $ZZ$ \\
\hline
$\oO_1$ & 64
& -1 & 1 & 1 & -1 & -1 & 1 & 1 & 1 & -1 & -1 & -1 & 1 & -1 & -1 & 1 \\
$\mathcal{O}_2$ & 576
& 0 & 1 & 0 & 0 & 0 & 0 & -1 & -1 & -1 & 0 & 0 & 0 & 1 & -1 & 1 \\
$\mathcal{O}_3$ & 288
& 0 & 0 & -1 & -1 & 0 & 0 & 0 & -1 & -1 & 0 & 1 & -1 & 1 & 0 & 0 \\
$\mathcal{O}_4$ & 288
& 0 & 0 & 0 & 0 & 0 & 0 & -1 & -1 & -1 & -1 & 1 & -1 & 1 & -1 & 1 \\
$\mathcal{O}_5$ & 192
& 0 & 0 & 0 & 0 & 0 & 0 &  1 &  1 &  1 &  1 & -1 & 1 &  1 &  1 & -1 \\
\hline
\end{tabular}
\caption{{Orbit representatives for the vertices of the $(2,3,2)$ nonsignaling polytope in expectation coordinates. Orbit $\oO_{1}$ consists of deterministic vertices. Orbits $\oO_{2}$ and $\oO_{3}$ can be viewed as a lifting of the Popescu-Rohrlich (PR) box \cite{popescu1994quantum} from the $(2,2,2)$ Bell scenario to additional measurement settings. Orbits $\oO_{4}$ and $\oO_{5}$ consist of max-weight vertices.}}
\label{tab:ns232_orbits_verbose}
\end{table}

We now focus on the locally closed operators $A_\Omega^\gamma$ and provide criteria for determining when they correspond to vertices of the locally closed polytope.
{To} this end, we introduce a partial order on the set of local pairs.
 Given local pairs $(\Omega_i,\gamma_i)$, where $i=1,2$, we write $(\Omega_1,\gamma_1) \preceq (\Omega_2,\gamma_2)$ if $\Omega_1\subseteq \Omega_2$ and $\gamma_2|_{\Omega_1}=\gamma_1$.
A local pair $(\Omega,\gamma)$ is called maximal if it is maximal with respect to the partial order $\preceq$.

\begin{Lemma}\label{lem:vertex maximal}
If $(\Omega,\gamma)$ is maximal
{then}
 $A_{\Omega}^\gamma$ is a vertex of {${\LC_n}$}.
\end{Lemma}
{\em{Proof.}}
Assume that the operator is not a vertex. That is, we can write
$$
A_{\Omega}^\gamma = \sum_{i=1}^k \lambda_i A_{\Omega_i}^{\gamma_i}
$$
where $\lambda_i \in \RR_{\geq 0}$, $\sum_i \lambda_i=1$, $k\geq 2$, {and the terms in the sum are distinct}.
{Since
\begin{align*}
(-1)^{\gamma(a)} &= \Tr(A_{\Omega}^{\gamma}T_a)\\
&= \sum_i \lambda_i \Tr(A_{\Omega_i}^{\gamma_i}T_a)\\
&= \sum_i \lambda_i (-1)^{\gamma_i(a)}
\end{align*}
for all $a\in \Omega$,} we have $(\Omega_i,\gamma_i)\succeq (\Omega,\gamma)$. Thus $(\Omega,\gamma)$ is not maximal.
\hfill$\square$\par
 \medskip

\begin{Corollary}\label{cor:max weight vertices}
The {deterministic vertices and the} max-weight vertices of $\LP_n$ are also vertices of $\LC_n$.
\end{Corollary}
\begin{proof}
{The deterministic vertices are defined by local value assignments on the whole space $E_n$, and hence are automatically maximal. For max-weight vertices, maximality follows from Lemma~\ref{lem:maximal vs bipartite}.}
\end{proof}

\section{Locally closed vertex simulators}
\label{sec:locally closed vertex simulators}

\subsection{{Update rules}}
  Next, we describe the update rules for locally closed operators. {The corresponding formulas are similar to the update rules for CNC operators; see~\cite{raussendorf2020phase,ipek2025phase}.}
Let $E_{n-1}^{(i)}$ denote the subspace of $E_n$ spanned by $x_j,z_j$ with $j\neq i$. A locally closed set $\Omega$ can be decomposed into pairwise disjoint subsets
\begin{equation}\label{eq:locally closed decomposition}
\Omega = \bigcup_{c\in \Span{x_i,z_i}} \Omega^{(i)}(c),
\end{equation}
where $\Omega^{(i)}(c)=\Omega\cap(c+E_{n-1}^{(i)})$. For simplicity of notation, we write $\Omega(c)$ when there is no danger of confusion.
Using this decomposition, for $c\in \set{x_i,y_i,z_i}$ we define a locally closed set in $E_{n-1}^{(i)}$ by
\[
\Omega_c=\Omega(0)\cup (c+\Omega(c)).
\]
Note that we can identify $E_{n-1}^{(i)}$ with $E_{n-1}$ by shifting the indices $x_j,z_j$ down for $j\geq i+1$. Under this identification, $c+\Omega(c)$ and $\Omega_c$ can be regarded as locally closed subsets of $E_{n-1}$.

 {Our computational model (MBPC) uses destructive {local} measurements. In this case the update rules are given below. Therein, $\Omega(0)$ and $\Omega_b$ are regarded as subsets of $E_{n-1}$.}

\begin{Lemma}\label{lem:update local CNC}
Let $b\in \set{x_i,y_i,z_i}$
and $(\Omega,\gamma)$ be a local pair.
\begin{enumerate}
\item If $b\in \Omega$ then
\begin{equation}\label{eq:b in Omega}
\Tr_i(\Pi_b^r A_\Omega^\gamma \Pi_b^r)  = \delta_{r,\gamma(b)} A_{\Omega(0)}^{\gamma|_{\Omega(0) }}.
\end{equation}
\item If $b\not\in \Omega$ then
\begin{equation}\label{eq:b not in Omega}
\Tr_i(\Pi_b^r A_\Omega^\gamma \Pi_b^r) = \frac{1}{2} A_{\Omega_b}^{\gamma_b^r}
\end{equation}
where
$\gamma_b^r:\Omega_b\to \ZZ_2$ is a local value assignment defined by
$$
\gamma_b^r(a) =
\left\lbrace
\begin{array}{ll}
\gamma(a) & a\in \Omega(0) \\
\gamma(a+b)+r & a\in b+\Omega(b).
\end{array}
\right.
$$
\end{enumerate}
\end{Lemma}
\begin{proof}
{For notational simplicity we will assume that the measurement is performed on the last qubit.} {Define the zero-extended Pauli coefficients $\chi: E_n \to \{0, \pm 1\}$ by
\[
  \chi(a) = \begin{cases}
    (-1)^{\gamma(a)} & a \in \Omega, \\
    0 & a \notin \Omega.
  \end{cases}
\]
Then we have
\[
  \Pi_b^r A_\Omega^\gamma \Pi_b^r = \bar{A} \otimes \Pi_b^r,
\]
where
\[
  \bar{A} = \frac{1}{2^n} \sum_{a \in E_{n-1}} \left( \chi(a) + (-1)^r \chi(a+b) \right) T_a.
\]
}

There are two cases to {consider}:
\begin{enumerate}
\item $b\in \Omega$: $\bar A=\zero$ if $\gamma(b)\neq r$. Otherwise,
$$
\begin{aligned}
\bar A  & = \frac{1}{2^{n-1}} \sum_{a\in \Omega\cap E_{n-1}} (-1)^{\gamma(a)} T_a
=   A_{\Omega\cap E_{n-1}}^{\gamma|_{\Omega \cap E_{n-1}}}.
\end{aligned}
$$
\item $b\notin \Omega$:
$$
\begin{aligned}
\bar A & = \frac{1}{2^{n-1}} \left(
\sum_{a\in \Omega\cap E_{n-1}} \frac{(-1)^{\gamma(a)}}{2} T_a
+
{\sum_{a\in (b+\Omega)\cap E_{n-1}}}\frac{(-1)^{\gamma(a+b)+r}}{2} T_a
\right)
= \frac{1}{2} A_{\Omega_b}^{\gamma_b^r}.
\end{aligned}
$$
\end{enumerate}
\end{proof}

{
\begin{Rem}\label{rem:non destructive update and cncs}
The update rules for the non-destructive case can be obtained by the same computation as in the proof of Lemma~\ref{lem:update local CNC}. For $\Omega\subseteq E_{n-1}$ and $b\in \{x_n,y_n,z_n\}$, we write $\langle \Omega,b \rangle$ for the subset given by the disjoint union of $\Omega$ and $\Omega+b$. For a local value assignment $\gamma:\Omega\to \ZZ_2$ and $r\in \ZZ_2$, we write $\gamma\star r$ for the local value assignment
\[
\gamma\star r(a) =
\begin{cases}
\gamma(a) & a\in \Omega, \\
\gamma(a)+r & a\in \Omega+b.
\end{cases}
\]
Then the update rules for non-destructive local Pauli measurements are given by
\[
\Pi_b^r A_\Omega^\gamma \Pi_b^r =
\begin{cases}
\delta_{r,\gamma(b)} A^{\gamma|_{\Omega(0)}\star r}_{\langle \Omega(0),b\rangle}, & {b \in \Omega}\\[1em]
\frac{1}{2} A_{\langle \Omega_b,b \rangle}^{\gamma_b^r\star r} & {b \notin \Omega}.
\end{cases}
\]
In particular, if $A_\Omega^\gamma$ is a CNC operator, then the operator obtained after a destructive or non-destructive local Pauli measurement remains a CNC operator up to normalization.
\end{Rem}
}


\subsection{{The deterministic vertex simulator}}
\label{sec:deterministic simulator}

The deterministic vertices described in Section~\ref{sec:deterministic-vertices} are of special interest for their role in defining the Bell polytope of local correlations. From the computational perspective, as we demonstrate below, they also admit an efficient classical simulation algorithm. Thus, distinct notions of classicality coming from both quantum foundations and computer science coincide within the setting of MBPC \cite{raussendorf2017contextuality}. In the case of non-adaptive computation it is known that deterministic vertices have limited computational power, being restricted to the deterministic computation of $n$-partite linear functions \cite{hoban2011generalized,raussendorf2013contextuality}.

Let us describe the update maps for both destructive and non-destructive
single-qubit Pauli measurements acting upon a deterministic vertex. By
Equation~\eqref{eq:deterministic}, the update maps reduce to the $8$-state
model in the single-qubit case. Consider first a non-destructive measurement of the Pauli $X$ operator. We have
\begin{eqnarray}
\Pi_x^s A^{tuv}\Pi_x^s = \delta_{t,s}\Pi_x^s,\notag
\end{eqnarray}
which can be decomposed as
\begin{eqnarray}
\Pi_x^s A^{tuv}\Pi_x^s
=
\delta_{t,s}
\left(
\frac{A^{tuv}+A^{t\bar u\bar v}}{2}
\right),
\label{eq:det-single-update}
\end{eqnarray}
where $\bar u=u+1$ and $\bar v=v+1$. Similar relations hold for measurements of the Pauli $Y$ and $Z$ operators. Thus the measurement outcome is deterministic, while the post-measurement vertex is a uniform mixture over the two deterministic vertices compatible with the measured value.

In the $n$-qubit non-destructive case, suppose for convenience that
$a_1\in E_n$ is a local Pauli supported on the first qubit. Then
\[
\Pi_{a_1}^s A_{E_n}^{\gamma}\Pi_{a_1}^s
=
\delta_{\gamma(a_1),s}
\left(
\frac{A_{E_n}^{\gamma}+A_{E_n}^{\bar\gamma}}{2}
\right),
\]
where $\bar\gamma$ agrees with $\gamma$ on the last $n-1$ qubits and is updated on the first qubit according to Eq.~\eqref{eq:det-single-update}. Hence the outcome is deterministically given by $s=\gamma(a_1)$, while the hidden deterministic vertex updates randomly to either $A_{E_n}^{\gamma}$ or $A_{E_n}^{\bar\gamma}$ with equal probability.

These update rules further simplify in the destructive case, since the
operator update is followed by a partial trace over the measured qubit. This yields
\[
\Tr_1\left(
\Pi_{a_1}^s A_{E_n}^{\gamma}\Pi_{a_1}^s
\right)
=
\delta_{\gamma(a_1),s}
A_{E_{n-1}}^{\gamma|_{n-1}},
\]
because $\gamma$ and $\bar\gamma$ agree after restriction to the last $n-1$ qubits. After a sequence of $k\leq n$ destructive measurements we obtain
\begin{eqnarray}
\Tr_{1\cdots k}\left(
\Pi_{a_k}^{s_k}\cdots \Pi_{a_1}^{s_1}
A_{E_n}^{\gamma}
\Pi_{a_1}^{s_1}\cdots \Pi_{a_k}^{s_k}
\right)
=
\left(
\prod_{i=1}^k \delta_{\gamma(a_i),s_i}
\right)
A_{E_{n-k}}^{\gamma|_{n-k}}.
\end{eqnarray}
Thus destructive measurements of deterministic vertices have deterministic
outcomes and deterministic updates.

With regard to reproducing the measurement statistics of deterministic
vertices there is a simple and efficient classical algorithm. See Algorithm~\ref{alg:deterministic-sampling}. We assemble the images $\gamma(a_i)$ for all $3n$ local Pauli elements into a vector
\[
v_{\gamma}=
(\gamma(x_1),\gamma(y_1),\gamma(z_1),\ldots,
\gamma(x_n),\gamma(y_n),\gamma(z_n))
\in \ZZ_2^{3n}.
\]
The simulator thus stores one deterministic value for each local Pauli measurement, using $3n$ bits of memory. At the $i$-th step, the
measurement label $a_i\in\{x_i,y_i,z_i\}$ specifies one of the three entries in the $i$-th block, so the outcome $r_i=\gamma(a_i)$ is obtained in constant time. The update consists only of incrementing the counter $i\mapsto i+1$. Hence a sequence of $n$ destructive local Pauli measurements can be simulated using $O(n)$ memory and $O(n)$ time.


\begin{algorithm}
\justifying
\begin{enumerate}[leftmargin=*]

\item For $i=1,\ldots,n$, given the destructive local Pauli measurement $a_i$ on the $i$-th qubit, output
\[
r_i=\gamma(a_i).
\]

\item Output $r=(r_1,\dots,r_n)$.
\end{enumerate}
\caption{\label{alg:deterministic-sampling}
Classical algorithm to sample from the outcome distribution of the deterministic vertex associated with a local value assignment $\gamma$ under a sequence $a_1,\ldots,a_n$ of destructive local Pauli measurements.
}
\end{algorithm}

{
Algorithm~\ref{alg:deterministic-sampling} can be viewed as a concrete instance of the ncHVM simulator of~\cite{raussendorf2017contextuality}. In their algorithm, a resource quantum state admitting an ncHVM description is first expanded as a convex decomposition of quantum states, i.e., an ensemble, each of which is represented non-negatively by one of the conjugate Wigner functions.
The algorithm then begins by sampling from the initial ensemble and proceeds with the simulation using the corresponding Wigner function. Their subroutine requires measurements that preserve the positivity of the chosen Wigner representation. Our Algorithm~\ref{alg:deterministic-sampling} provides a concrete model for this approach. Quantum states $\rho$ that admit an ncHVM with respect to local Pauli measurements are precisely those contained in the deterministic polytope. The initial step in our algorithm is a convex decomposition of $\rho$ in terms of the vertices of $\on{DP}_n$. Recall from Section~\ref{sec:deterministic-vertices} that these vertices are partitioned into conjugate phase-point operators. The update rules of Algorithm~\ref{alg:deterministic-sampling} are the destructive counterparts of the Wigner simulation used in~\cite{raussendorf2017contextuality}.
}

\subsection{The max-weight vertex simulator}

Here we consider the classical simulation of the max-weight vertices
introduced in Section~\ref{sec:power of correlations}.
Let $\gamma:\ZZ_3^n\to\ZZ_2$ be a
{non-bipartite-linear (NBL)} function.
{Using the identification and notation introduced there, the
corresponding max-weight operator is}
\[
{A_{E_n^\circ}^{\gamma}}
=
\frac{1}{2^n}
\left(I+\sum_{a\in E_n^\circ}(-1)^{{\gamma(a)}}T_a\right).
\]

{Since the nonidentity Pauli terms of
$A_{E_n^\circ}^{\gamma}$ all have weight $n$, we have}
\[
\Tr\!\left(A_{E_n^\circ}^{\gamma}T_a\right)=0
\]
for every nonzero $a\in E_n\setminus E_n^\circ$. In particular, for any single local Pauli measurement $\Pi_{a_i}^r$,
\[
\Tr\!\left(A_{E_n^\circ}^{\gamma}\Pi_{a_i}^r\right)=\frac12,
\]
so each individual measurement outcome is uniformly random.

{Nevertheless, the outcomes of a full set of $n$ local Pauli measurements, one on each qubit, are strongly correlated. If $a_1,\dots,a_n$ are local Pauli elements and
\[
a=a_1+\cdots+a_n\in E_n^\circ,
\]
then
\[
\Tr\!\left(A_{E_n^\circ}^{\gamma}\Pi_{a_1}^{r_1}\cdots \Pi_{a_n}^{r_n}\right)
=
\begin{cases}
2^{1-n}, & {\gamma}(a)=r_1\oplus\cdots\oplus r_n,\\
0, & \text{otherwise}.
\end{cases}
\]
Thus the joint distribution is uniform over the $2^{n-1}$ outcome strings whose parity is fixed by $\gamma(a)$. This immediately yields a classical sampling algorithm; see Algorithm~\ref{alg:max-weight-sampling-short}.}

In general, max-weight vertices do not admit an efficient
representation. To specify such an operator, one needs to choose
a $\pm 1$ coefficient for each of the $3^n$
{Pauli operators of weight $n$}.
Nevertheless, the non-adaptive case has a simpler structure.
{This case is also computationally interesting because of
its connection with instantaneous quantum computation (IQP).}
{Efficient exact classical sampling of general IQP circuits
would imply a collapse of the polynomial hierarchy, and is
therefore considered highly unlikely}~\cite{bremner2011classical}.
An IQP circuit is initialized in the state $\ket{+}^{\otimes n}$,
transformed by gates diagonal in the computational basis, and
terminates with non-adaptive Pauli $X$ measurements; see, for
instance,~\cite{jozsa2024iqp}.
{For the reduced resource states of
Theorem~\ref{thm:t qubit resource}, non-adaptive MBPC with
$U=\varnothing$ can be written as a restricted family of
IQP circuits.}
This is done by grouping together the entangling gate $E(G)$
of the initial graph state, the $T$ gates acting on the magic
part of the resource state, and the $S^\dagger$ gates arising
from the decomposition $M(\pi/2)=M(0)S^\dagger$ whenever a
$Y$ measurement is applied.
The computation then consists of applying these diagonal gates
to $\ket{+}^{\otimes n}$, followed by non-adaptive Pauli $X$
measurements.
{For an individual max-weight operator,
Algorithm~\ref{alg:max-weight-sampling-short} is efficient for
a fixed non-adaptive measurement context, provided that the
corresponding value of $\gamma$ is available.}
It only requires uniformly sampling $n-1$ bits and using the
stored value of the local assignment at the max-weight element
determined by the sum of the chosen local measurements.

\subsubsection{NBL functions with efficient description}

{
For general adaptive computations, Algorithm~\ref{alg:max-weight-sampling-short} is not typically efficient, since an arbitrary sequence of local Pauli measurements may occur and evaluation of a general NBL function from its full specification can require exponential time in $n$.
We therefore restrict attention to specific families {$\operatorname{NBL}^{\text{eff}}$} that admit an efficient, polynomial-size description in $n$.

For $n\in\NN$, fix an integer $p$ with $2\le p\le n$, and let $u=(u_1,\dots,u_n)\in\ZZ_3^n$. We exclude the case $p=1$ since the resulting function is bipartite linear. Define
\begin{equation}
\label{eq:gamma p efficient}
\gamma^{(p)}(u)
:=
\left(
\sum_{1\le i_1<\cdots<i_p\le n}
u_{i_1}u_{i_2}\cdots u_{i_p}
\right)\bmod 2,
\end{equation}
where each product $u_{i_1}\cdots u_{i_p}$ is first evaluated in $\ZZ_3$, and the final sum is then reduced modulo $2$.

To see that this description is efficient, note that there are
\[
\binom{n}{p}=O(n^p)
\]
products in the sum. Each product requires $p-1$ multiplications in the fixed finite ring $\ZZ_3$, and hence can be evaluated in $O(p)$ time. Since reduction modulo $2$ can be performed incrementally during the summation, the total time required to evaluate $\gamma^{(p)}$ is
\[
O\!\left(p\binom{n}{p}\right)=O(n^p)
\]
for fixed $p$.
}

\paragraph{Example}%
Consider the quadratic case $p=2$, for which
\[
\gamma^{(2)}(u)=\sum_{i<j}u_i u_j \mod 2.
\]
There are $n(n-1)/2$ terms, each involving one multiplication in $\ZZ_3$. For $n=2$, this reduces to
\[
\gamma^{(2)}(u_1,u_2)=u_1u_2 \pmod 2,
\]
where the product is first evaluated in $\ZZ_3$ and the result is then reduced modulo $2$. Its truth table is
\[
\begin{array}{c|ccc}
u_{1}/u_{2} & 0 & 1 & 2\\
\hline
0 & 0 & 0 & 0\\
1 & 0 & 1 & 0\\
2 & 0 & 0 & 1
\end{array}
\]
where the rows are indexed by $u_1$ and the columns by $u_2$. This is precisely the representative for orbit $\oO_{5}$ in Table~\ref{tab:ns232_orbits_verbose}.
 {The orbit $\oO_{4}$ also corresponds to a quadratic NBL function that is not symmetry-equivalent to $\gamma^{(2)}$.}

\begin{algorithm}
\justifying
\begin{enumerate}[leftmargin=*]
\item For the local Pauli measurements $a_1,\dots,a_{n-1}$, sample outcomes $r_1,\dots,r_{n-1}\in \ZZ_2$\\
independently and uniformly at random.

{
\item For $a_n$, set
\[
r_n=\gamma(a)\oplus r_{1}\oplus \cdots \oplus r_{n-1},
\]
where
\[
a=a_1+\cdots+a_n\in E_n^\circ.
\]}

\item Output $r=(r_1,\dots,r_n)$.
\end{enumerate}
\caption{\label{alg:max-weight-sampling-short}
Classical algorithm to sample from the joint outcome distribution of the max-weight operator $A_{E_n^\circ}^{\gamma}$ under a sequence $a_1,\dots,a_n$ of local Pauli measurements with $a=a_1+\cdots+a_n\in E_n^\circ$.
}
\end{algorithm}

\section{{Local robustness and efficiency}}
\label{sec:local-robustness-efficiency}

Algorithm~\ref{alg:Algorithm} provides a simulation of any quantum computation as a probabilistic classical computation. The key feature of the polyhedral method is that, at each quantum process implementing a step of the computation, the preservation property described in Definition~\ref{def:preserved} guarantees that the transformed vertex operator lies in the polytope determined by the convex hull of the operators used at the next step. The resulting sampling algorithm is exact, in the sense that it produces samples from the Born rule probabilities of the original computation.

To run this algorithm in practice, for instance for the local Pauli simulator, one needs to know all the vertices of $\LP_n$, or at least those relevant to the initial state and to the regions that can be reached under the measurements, in order to determine the classical update rules. A way around this problem is to restrict to a subpolytope of $\LP_n$ whose vertices have a known description and whose update rules are known and typically efficient. The prototypical model for this approach is the CNC simulator of~\cite{ipek2025phase}, together with earlier simulators based on the stabilizer polytope~\cite{gottesman1998heisenberg,aaronson2004improved} and the Wigner polytope~\cite{veitch2012negative}.

\subsection{{Local robustness measure}}
\label{sec:local robustness measure}

{In general, there are two components to the complexity analysis of a polyhedral simulator:
\begin{itemize}
\item \emph{Sample complexity}: the number of independent runs required to achieve a desired precision. In the Monte Carlo quasiprobability-estimation framework, this is governed by $\mathfrak{R}_{{P}}(\rho)^2$.

\item \emph{Per-sample complexity}: the cost of each run, consisting of sampling from the initial distribution and from the distributions associated with the update maps.
\end{itemize}
The robustness measure $\mathfrak{R}_{{P}}(\rho)$ is a key quantity in the analysis of the sample complexity (Definition \ref{def:robustness}). It arises by relaxing from probability distributions to quasi-probability distributions:}
{For simulating quantum computations initialized at states that do not necessarily lie in such a restricted polytope, one can relax the requirement that the initial state lie in the convex hull of the operators, i.e., within the polytope, to the requirement that it lie in their affine hull. In effect, this means that the initial decomposition
\[
\rho = \sum_{\alpha \in V(P)} p_\rho(\alpha) A_\alpha,\quad p_\rho(\alpha)\in \RR,\;\sum_{\alpha} p_{\rho}(\alpha)=1,
\]
may contain negative coefficients, and thus $p_\rho$ forms a \emph{quasi-probability} distribution.}

Using the methods of~\cite{pashayan2015estimating}, this quasi-probability distribution $p:V(P)\to \RR$ can be converted into a probability distribution $\tilde p:V(P)\to \RR_{\geq 0}$ by setting
\[
\tilde p(\alpha)=\frac{|p(\alpha)|}{\norm{p}_1},
\qquad
\norm{p}_1=\sum_{\alpha}|p(\alpha)|.
\]
Here $P$ is the relevant polytope, for example the CNC polytope, defined as the convex hull of CNC operators, or the stabilizer polytope, defined as the convex hull of stabilizer states. An optimal quasi-probability distribution can then be chosen and transformed in this way to obtain a probability distribution that can be used to initialize the simulation algorithm. {For a fixed target outcome $s^*$, the output $(s,w)$ of
Algorithm~\ref{alg:estimation} gives the unbiased estimator
$w\,\delta_{s,s^*}$ of its Born rule probability;
see also~\cite{okay2025polyhedral}.}

\begin{algorithm}
\justifying
\begin{enumerate}[leftmargin=*]
\item Sample from the probability distribution
{$\tilde p_\rho(\alpha)=|p_\rho(\alpha)|/\norm{p_\rho}_1$}
to obtain the initial vertex label $\alpha_0\in V(P_0)$.
{Set $w=\norm{p_\rho}_1\operatorname{sgn}(p_\rho(\alpha_0))$.}

\item For $i=1,\ldots,N$, sample from the corresponding update map
$q_{\alpha_{i-1}}^{(i)}$ to obtain a new vertex label
$\alpha_i\in V(P_i)$ and an outcome $s_i\in\Sigma_i$.
{Record} $s_i$ and update the vertex label
$\alpha_{i-1}\mapsto\alpha_i$.

\item {Output $(s,w)$, where $s=(s_1,\ldots,s_N)$.}
\end{enumerate}
\caption{\label{alg:estimation}
Classical algorithm based on the polytopes $P_0,\ldots,P_N$ for
{estimating the outcome probabilities} of a quantum computation
consisting of a sequence $\Phi_1,\ldots,\Phi_N$ of adaptive instruments,
where each $\Phi_i$ preserves the pair $(P_{i-1},P_i)$,
initialized at {$\rho$ with an affine decomposition $p_\rho$
over $V(P_0)$}.
}
\end{algorithm}

The resulting {sample} overhead is proportional to
{the square of} the measure defined next.

{
\begin{Definition}\label{def:robustness}
Let $P\subseteq \on{Herm}_1(\hH)$ be a polytope. The \emph{robustness measure} is defined by
\begin{equation}\label{eq:robustness}
\mathfrak{R}_{P}(\rho) = \min_p \left\lbrace
\norm{p}_1 : \rho = \sum_{\alpha\in V(P)} p(\alpha) A_\alpha,\; p(\alpha)\in \RR,\;\sum_\alpha p(\alpha)=1
\right\rbrace,
\end{equation}
where $V(P)$ denotes the vertex set of $P$.
\end{Definition}
}

{\begin{Prop}\label{pro:robustness properties}
The robustness measure satisfies the following properties. Let $A,B\in \on{Herm}_1(\hH)$, and let $\rho$ be a quantum state.
\begin{enumerate}
\item Faithfulness: $\mathfrak{R}_P(A)\geq 1$ and $\mathfrak{R}_P(A)=1$ if and only if $A\in P$.
\item Convexity: For $\{A_\alpha\in \on{Herm}_1(\hH)\}$ and $\{\lambda_\alpha \in \RR : \sum_\alpha \lambda_\alpha=1\}$, where $\alpha=1,\dots,n$,
\[
\mathfrak{R}_P\left(\sum_\alpha \lambda_\alpha A_\alpha\right) \leq \sum_\alpha |\lambda_\alpha|\mathfrak{R}_P(A_\alpha).
\]
\item Submultiplicativity: If $P$ is closed under tensor products, then
\[
\mathfrak{R}_P(A\otimes B) \leq \mathfrak{R}_P(A)\mathfrak{R}_P(B).
\]
\item Monotonicity: Let $Q\subseteq \on{Herm}_1(\kK)$ be a polytope, and let $\Phi$ be an instrument that preserves the pair $(P,Q)$. Writing $\PP(s)=\Tr(\Phi^s(\rho))$, we have
\[
\sum_{s:\, \PP(s)>0} \PP(s)\,
\mathfrak{R}_Q\left(\frac{\Phi^s(\rho)}{\PP(s)}\right)
\leq
\mathfrak{R}_P(\rho).
\]
\end{enumerate}
\end{Prop}
\begin{proof}
The first property is standard for robustness measures, including the robustness of magic; see, for instance, \cite{howard2017application}. The second property is also standard and follows by choosing an optimal decomposition for the robustness measure with respect to $P$. For the third property, we use Proposition~\ref{pro:tensor-vertices-gpt}. By this result, when the polytope is closed under tensor products, then for vertices $A_{\alpha},A_{\beta}\in P$, their product $A_\alpha\otimes A_\beta$ is also a vertex of $P$. The proof follows from this property by considering optimal expansions for $A$ and $B$.
For the last property, let
$
\rho=\sum_\alpha p(\alpha)A_\alpha
$
be an optimal decomposition with respect to $P$. Using the update map $q_\alpha$ for the instrument $\Phi$, we have
$
\Phi^s(A_\alpha)=\sum_{\beta}q_\alpha(\beta,s)A_\beta.
$
Therefore
$
\Phi^s(\rho)=\sum_{\alpha,\beta}p(\alpha)q_\alpha(\beta,s)A_\beta$.
Summing over all possible outcomes $s$ gives
\[
\sum_{s:\PP(s)>0} \PP(s)\,\mathfrak{R}_Q\left(\frac{\Phi^s(\rho)}{\PP(s)}\right)
\leq
\sum_{\alpha,\beta,s}|p(\alpha)|q_\alpha(\beta,s)
=
\sum_\alpha |p(\alpha)|
=
\mathfrak{R}_P(\rho),
\]
where we used $\sum_{\beta,s}q_\alpha(\beta,s)=1$.
\end{proof}

The main examples of robustness measures come from stabilizer theory and its extension based on CNC phase space. The relevant polytopes are the CNC polytope $\on{CP}_n$, given by the convex hull of the CNC operators, and the stabilizer polytope $\on{SP}_n$.
We write $\mathfrak{R}_{\on{C}}(\rho)$ for the robustness measure associated with the CNC polytope, also called \emph{phase-space robustness}~\cite{raussendorf2020phase}. Similarly, we write $\mathfrak{R}_{\on{S}}(\rho)$ for the robustness measure associated with the stabilizer polytope, also known as the \emph{robustness of magic}~\cite{howard2017application}. The latter satisfies all the properties listed in Proposition~\ref{pro:robustness properties}, while the former satisfies all except submultiplicativity, since the CNC polytope is not tensor closed. We have the polytope inclusion
$
\on{SP}_n \subseteq \on{CP}_n,
$
and the comparison of robustness measures follows:
\begin{equation}\label{eq:robustness comparison S vs C}
 \mathfrak{R}_{\on{C}}(\rho) \leq \mathfrak{R}_{\on{S}}(\rho).
\end{equation}
}

{The locally closed operators in the local Pauli polytope give rise to a new simulation method {and to a corresponding robustness measure}.}
Let $V^{\eff}_n$ consist of locally closed operators that have
\begin{itemize}
\item an efficient classical representation in $n$, and
\item efficient update rules under single-qubit Pauli measurements.
\end{itemize}
Observe that, {as a consequence of Theorem~\ref{thm:tensor},} the set $V^{\eff}_n$ is closed under tensor products; that is, if $A\in V^{\eff}_k$ and
$B\in V^{\eff}_m$, then $A\otimes B\in V^{\eff}_{k+m}$. The efficiency of the representation of the tensor product is clear, and the efficiency of the update rules follows from the locality of the measurements.
Important classes of operators contained in this set include
\begin{itemize}
\item the CNC operators of $\Lambda_n$,
\item the deterministic vertices of $\LP_n$,
\item {max-weight vertices specified by $\operatorname{NBL}^{\text{eff}}$ functions,}
\item the vertices of $\LP_2$.
\end{itemize}
{We can enlarge the set of efficiently representable operators by including their} tensor products.

{
\begin{Definition}\label{def:local and deterministic robustness}
We write $\LC_n^{\eff}$ for the polytope of
\emph{efficiently representable locally closed operators}, defined
as the convex hull of $V_n^{\eff}$.
For a quantum state $\rho$, the corresponding robustness measure,
\emph{local robustness}, will be denoted by $\mathfrak{R}_L(\rho)$.
The deterministic robustness $\mathfrak{R}_D(\rho)$ is defined
with respect to the deterministic polytope $\on{DP}_n$.
\end{Definition}
}

 These robustness measures satisfy all the properties listed in Proposition~\ref{pro:robustness properties}. The inclusions $\on{DP}_n\subseteq \on{LC}_n^{\on{eff}}$ and $\on{CP}_n\subseteq \on{LC}_n^{\on{eff}}$ imply the following comparisons:
\begin{equation}\label{eq:robustness comparison D vs L vs C}
\mathfrak{R}_D(\rho) \geq  \mathfrak{R}_L(\rho) \leq \mathfrak{R}_C(\rho).
\end{equation}
}

{Typically, we choose a subset of locally closed operators that have efficient update rules. Thus, the update part of the per-sample complexity is efficient for the simulators considered here. However, the complexity of sampling from an optimal initial decomposition is not known in general. Consequently, an improvement in the robustness measure directly gives an improvement in sample complexity, but does not by itself imply an improvement in the total computational complexity of the simulation. When the initial distribution is also efficiently samplable, the only source of inefficiency comes from the sample complexity. In particular, in this case $\mathfrak{R}_{{P}}(\rho)=1$ implies an efficient classical simulation. The canonical efficient simulator that naturally arises from the locally closed operators is the deterministic-vertex simulator of Section~\ref{sec:deterministic simulator}.
Any quantum state that lies inside this polytope and admits an efficiently samplable convex decomposition can be efficiently simulated.}

 {
For the MBPC model, we can observe two immediate efficiently simulatable limiting cases.
Consider the resource state $\ket{\psi_{G,U}}=E(G)T^{\otimes U}\ket{+}^{\otimes n}$ in the cases where it (1) has no \emph{entanglement}, i.e., $E(G)=\one$, and (2) has no \emph{magic}, i.e., $U=\emptyset$.
In the first case, the resource state is a product magic state. Expanding each single-qubit state inside the deterministic polytope and taking their tensor product gives a convex decomposition inside the deterministic polytope. Thus, the deterministic robustness and the local robustness measures of this state are both $1$. On the other hand, the second case is a graph-state, or cluster-state, resource. Any stabilizer state is contained in the convex hull of the CNC operators, which is contained in the efficiently representable sector of the locally closed polytope. Thus, the local robustness is $1$. We conclude that, in our model, both magic and entanglement need to be present for quantum computational advantage.
}

\subsection{Reduced resource states}
\label{sec:reduced-resource}

The standard conversion from a circuit to MBQC replaces the $H$ and $T$ gates with {their corresponding} measurement patterns. This typically has the drawback of increasing the number of qubits. However, using the theory of graph states, it is possible to reduce the number of qubits to the number of $T$ gates in the original circuit. Such a reduction is also important when comparing with quantum computations described in the QCM model.



\begin{figure}[htbp]
\centering

\begin{minipage}[t]{0.49\textwidth}
\centering
\[
\begin{quantikz}
\lstick{$\ket{\psi}$} & \ctrl{1} & \meter{X} \vcw{1} &  \\
\lstick{$\ket{+}$}    & \control{} & \gate{X}        & \rstick{$H\ket{\psi}$}\qw
\end{quantikz}
\]
\end{minipage}
\hfill
\begin{minipage}[t]{0.49\textwidth}
\centering
\[
 \begin{quantikz}
\lstick{$\ket{\psi}$} & \ctrl{1} & \gate{T}  & \meter{X} \vcw{1} &  \\
\lstick{$\ket{+}$}    & \control{} & \qw & \gate{X}        & \rstick{$J(\pi/4)\ket{\psi}$}\qw
\end{quantikz}
\]
\end{minipage}

\caption{Gadgets realizing $H$ and $J(\pi/4)$.}
\label{fig:gadgets}
\end{figure}

{
\begin{Theorem}\label{thm:t qubit resource}
Any $n$-qubit quantum circuit over the gate set $\{H_i,S_i,T_i,E_{ij}\}$ with $t$ $T$ gates can be efficiently converted into a measurement-based Pauli computation with {the reduced resource state}
\[
{\ket{\tilde \psi_{G,U}}} = T^{\otimes t} H^{\otimes U} E(G) \ket{+}^{\otimes t},
\]
where $G$ is a graph with no edges between nodes in $U$. The computation proceeds by adaptive Pauli $X/Y$ measurements and reproduces the output distribution of the original circuit up to efficient classical post-processing.
\end{Theorem}
\begin{proof}
Consider a quantum circuit initialized at $\ket{+}^{\otimes n}$, acted on by the unitary $U$, and terminating with Pauli $X$ measurements on all qubits, where $U$ has a decomposition over the universal gate set with $t$ $T$ gates and $h$ $H$ gates. Each $T$ gate can be decomposed further as $HJ(\pi/4)$. Therefore, we can express $U$ as a circuit over the gate set $\{H_i,J_i(\pi/4),S_i,E_{ij}\}$. Then $H_i$ can be implemented by the $J(0)$ pattern, introducing {$h+t$} ancilla qubits in the state $\ket{+}$, and similarly $J(\pi/4)$ introduces $t$ ancilla qubits in the state $\ket{+}$.
See Fig.~\ref{fig:gadgets} for the circuit description of these gadgets.

The $S$ gates in the decomposition of $U$ can be moved past the measurements, replacing an $M(\alpha)$ measurement with an {$M(\alpha-\pi/2)$} measurement. The resulting measurement pattern can be put into standard form, consisting of state preparation, entangling operations, and adaptive measurements in the $XY$ plane. Each $T$ gate introduces a non-Pauli measurement, namely an $M(-\pi/4)$ measurement. Thus, we arrive at an MBQC that begins with a graph state $E(G)\ket{+}^{\otimes(n+t+h)}$ and proceeds by an adaptive sequence of measurements $M(\alpha)$, where $\alpha\in \{0,\pm\pi/4,\pi/2\}$. The angles $0$ and $\pi/2$ correspond to Pauli $X$ and $Y$ measurements, respectively.

Next, we reduce the number of qubits by performing the Pauli measurements. The transformation of a graph state under single-qubit Pauli measurements is well known; see, for instance, \cite[Proposition 7]{hein2006entanglement}. Up to global phases and Pauli operators, the effect of local measurements is as follows:

\begin{itemize}
\item A $Z$-measurement on qubit $i$ transforms the graph by omitting this qubit.

\item A $Y$-measurement on qubit $i$ transforms the graph by local complementation at node $i$, followed by omitting node $i$, and produces $S$ gates on the neighbors of $i$ for either outcome $s=0,1$.

\item An $X$-measurement on qubit $i${, provided that $i$ has at
least one neighbor,} transforms the graph by local complementation at {an arbitrarily chosen} neighbor $i_0$ of $i$ followed by local complementation at $i$, omitting node $i$, and a final local complementation at $i_0$.
It produces an $H$ gate on qubit $i_0$ for either outcome $s=0,1$.
{If $i$ has no neighbors, the measurement
outcome is deterministically $+1$, and node $i$ is simply omitted.}
\end{itemize}
In summary, up to Pauli corrections, the measurement of a single-qubit Pauli operator produces $S$ and $H$ gates on the prescribed qubits. If these qubits are later measured in a Pauli basis, we can move these Clifford gates past the measurements and modify the measurement operators accordingly. Separating each $M(-\pi/4)$ measurement as the composite of a $T$ gate followed by an $X$ measurement, and commuting the remaining diagonal gates past the measurements, brings the computation into the following form:
\[
M_1(\alpha_1)M_2(\alpha_2) \cdots M_t(\alpha_t)T^{\otimes t} V_1V_2\cdots V_t E(\bar G) \ket{+}^{\otimes t},
\]
where the $V_i$ are local Clifford operators, and the measurement sequence is adaptive with angles $\alpha_i\in \{0,\pi/2\}$, followed by classical post-processing. This form of computation is also known as the \emph{measurement-based quantum computation on local Clifford decorated graph states} (LC-MBQC) model~\cite{backens2021there}. The local Clifford operators and the final graph $\bar G$ do not depend on the outcomes of the Pauli measurements.

A final reduction comes from the \emph{reduced graph states} of~\cite{elliott2008graphical}. Any stabilizer state, in particular $V_1V_2\cdots V_t E(\bar G) \ket{+}^{\otimes t}$, can be expressed as
\[
\prod_{i=1}^t H^{s_i}S^{r_i}Z^{u_i} E(G') \ket{+}^{\otimes t}
\]
for some graph $G'$. Moreover, for $U=\{ i : s_i=1\}$, we have $r_i=0$ if $i\in U$, and $G'$ has no edges between nodes in $U$. If $u_i=1$, the Pauli $Z$ operator can be moved past the $T$ gate and the measurements. Its effect can be taken into account by classical post-processing and/or by swapping measurements. Similarly, if $r_i=1$, the $S$ gate can be moved past the measurements and swaps them.

All steps in the construction are efficient. The translation of the original circuit into measurement patterns is linear in the number of gates. The reductions induced by Pauli measurements are implemented by local complementations and vertex deletions on graphs, which can be carried out in polynomial time. Finally, the reduced graph-state normal form of~\cite{elliott2008graphical} is constructive and can be found efficiently. Hence the graph $G$, the subset $U$, the adaptive Pauli measurement sequence, and the required classical post-processing can all be obtained efficiently from the original circuit.
\end{proof}

\begin{Ex}\label{ex:tht}
Consider the circuit obtained by measuring $THT\ket{+}$ in the $X$ basis. Rewriting $T=HJ(\pi/4)$, we can use the measurement patterns for the composite unitary $HJ(\pi/4)J(\pi/4)$, whose standard form is given by
\[
M_4(s_2\pi) M_3(s_1\pi) M_2(-(-1)^{s_1}\pi/4) M_1(-\pi/4) E_{12}E_{23}E_{34}.
\]
Up to flipping measurement outcomes, the measurements on qubits $3$ and $4$ are Pauli $X$ measurements. Performing the $X_4$ measurement removes qubit $4$ and the edge between qubits $2$ and $3$, and inserts an $H$ gate immediately before the $X_3$ measurement, up to Pauli operators and global phases. Moving this gate past the measurement converts it into a $Z_3$ measurement. Performing this measurement as well leaves the two-qubit resource state $T^{\otimes 2}E_{12}\ket{++}$. The computation then proceeds by first measuring $X_1$ and, depending on the outcome, measuring $X_2$ if $s_1=0$ or $Y_2$ if $s_1=1$.
\end{Ex}
}

\subsection{Comparison of {robustness measures}}
\label{sec:comparison of efficiency}

We can compare the efficiency of our simulation with existing approaches in the literature, mainly CNC simulation~\cite{raussendorf2020phase,ipek2025phase} and stabilizer simulation~\cite{howard2017application}.
{Equations~\eqref{eq:robustness comparison S vs C}--\eqref{eq:robustness comparison D vs L vs C} give}
\begin{equation}\label{eq:robustness comparison}
\mathfrak{R}_{\on{L}}(\rho) \leq \mathfrak{R}_{\on{C}}(\rho) \leq \mathfrak{R}_{\on{S}}(\rho).
\end{equation}
{Therefore, the {sample} complexity decreases as one passes from stabilizer simulation to CNC simulation and then to its local version.}



{We wish to establish a direct comparison between simulators. It is well-known that we can pass using stabilizer operations from a Clifford+$T$ circuit $C$ with $t$ $T$-gates to its magic-state injected variant and then finally to the corresponding PBC with resource state $\ket{T}^{\otimes t}$; see e.g., \cite{bravyi2016trading,yoganathan2019quantum}. {We will denote the resulting circuit by $C_{\on{PBC}}$.}
 On the other hand, the same circuit can be converted into an MBPC by replacing each $H$, $T$, and controlled-$Z$ gate with the corresponding pattern in the measurement calculus \cite{danos2007measurement}. Theorem~\ref{thm:t qubit resource} constructs the corresponding reduced $t$-qubit resource state for MBPC. {The resulting circuit will be denoted by $C_{\on{MBPC}}$.}}

{
\begin{Prop}\label{cor:direct comparison of robustness}
The $t$-qubit MBPC reduced resource state satisfies
\[
\mathfrak{R}_L( \ket{\tilde\psi_{G,U}}\bra{\tilde\psi_{G,U}} )
\leq
\mathfrak{R}_{C}((\ket{T}\bra{T})^{\otimes t}).
\]
Therefore, the {sample} complexity of simulating $C_{\on{MBPC}}$ using efficiently representable locally closed operators is at most that of simulating $C_{\on{PBC}}$ using CNC operators.
\end{Prop}
}
{
\begin{proof}
The reduced resource state $\ket{\tilde \psi_{G,U}}$ can be prepared from the product magic state $\ket{T}^{\otimes t}$ as follows. First tensor with the stabilizer state
$
\ket{\varphi}=H^{\otimes U}E(G)\ket{+}^{\otimes t}.
$
Then use the standard $T$-gadgets to implement the gates $T^{\otimes t}$ on $\ket{\varphi}$. These gadgets use only Clifford operations, adaptive local Pauli measurements, and outcome-dependent Clifford corrections.

It is known that tensoring a CNC vertex with a stabilizer state gives a CNC vertex~\cite{okay2021extremal}. Moreover, the CNC polytope is preserved by Clifford operations and destructive local Pauli measurements; see Remark~\ref{rem:non destructive update and cncs}. Therefore, the instruments implementing this preparation procedure preserve the CNC polytope. These instruments are deterministic on the input $(\ket{T}\bra{T})^{\otimes t}\otimes \ket{\varphi}\bra{\varphi}$, in the sense that the normalized post-measurement state is independent of the measurement outcome. Hence the averaged monotonicity statement of Proposition~\ref{pro:robustness properties} gives
\[
\mathfrak{R}_C(\ket{\tilde \psi_{G,U}}\bra{\tilde \psi_{G,U}})
\leq
\mathfrak{R}_C((\ket{T}\bra{T})^{\otimes t}\otimes \ket{\varphi}\bra{\varphi}).
\]
Since tensoring with a stabilizer state preserves CNC decompositions, we have
$
\mathfrak{R}_C((\ket{T}\bra{T})^{\otimes t}\otimes \ket{\varphi}\bra{\varphi})
\leq
\mathfrak{R}_C((\ket{T}\bra{T})^{\otimes t})$.
Thus
$
\mathfrak{R}_C(\ket{\tilde \psi_{G,U}}\bra{\tilde \psi_{G,U}})
\leq
\mathfrak{R}_C((\ket{T}\bra{T})^{\otimes t})$.
Combining this with the inclusion $\on{CP}_n\subseteq \on{LC}_n^{\on{eff}}$, equivalently with the right-hand side of Eq.~\eqref{eq:robustness comparison D vs L vs C}, gives the desired inequality.
\end{proof}
}

\begin{table}[t]
\centering
\small
\setlength{\tabcolsep}{4.5pt}
\begin{tabular}{cclrrrr}
\hline
$n$ & $G$ & $U$ & $\mathfrak R_S$ & $\mathfrak R_C$
& $\mathfrak R_{ L_0}$ & $\mathfrak R_{L_3}$ \\
\hline
2 & $P_2$ & $\varnothing$ & 1.748 & 1.000 & 1.207 & 1.000 \\
2 & $P_2$ & $\{1\}$ & 1.000 & 1.000 & 1.000 & 1.000 \\\hline
3 & $P_3$ & $\varnothing$ & 2.219 & 1.283 & 1.561 & {1.155} \\
3 & $P_3$ & $\{1\}$ & 1.414 & 1.000 & 1.707 & 1.000 \\
3 & $P_3$ & $\{2\}$ & 2.200 & 1.000 & 1.500 & 1.000 \\
3 & $P_3$ & $\{1,3\}$ & 1.414 & 1.000 & 1.414 & 1.000 \\\hline
3 & $K_3$ & $\varnothing$ & 2.219 & 1.283 & 1.737 & {1.136} \\
3 & $K_3$ & $\{1\}$ & 2.200 & 1.000 & 1.500 & 1.000 \\\hline
\end{tabular}
\caption{Complete robustness comparison for the inequivalent connected
decorated graphs on two and three vertices. We denote the path and completely connected graphs on $n$ vertices by $P_{n}$ and $K_{n}$, respectively.}
\label{tab:robustness-exhaustive-small}
\label{tab:comparison robustness}
\end{table}

\begin{table}[t]
\centering
\renewcommand{\arraystretch}{1.35}
\newcommand{\twograph}[2]{\begin{tikzpicture}[baseline=-0.5ex,scale=0.72]\coordinate (v1) at (-0.62,0);\coordinate (v2) at (0.62,0);\draw[line width=0.45pt] (v1)--(v2);\node[circle,draw,fill=#1,minimum size=3.2mm,inner sep=0pt,font=\tiny] at (v1) {1};\node[circle,draw,fill=#2,minimum size=3.2mm,inner sep=0pt,font=\tiny] at (v2) {2};\end{tikzpicture}}
\newcommand{\threegraph}[4]{\begin{tikzpicture}[baseline=-0.5ex,scale=0.72]\coordinate (v1) at (-0.72,-0.4);\coordinate (v2) at (0,0.48);\coordinate (v3) at (0.72,-0.4);\draw[line width=0.45pt] #1;\node[circle,draw,fill=#2,minimum size=3.2mm,inner sep=0pt,font=\tiny] at (v1) {1};\node[circle,draw,fill=#3,minimum size=3.2mm,inner sep=0pt,font=\tiny] at (v2) {2};\node[circle,draw,fill=#4,minimum size=3.2mm,inner sep=0pt,font=\tiny] at (v3) {3};\end{tikzpicture}}
\newcommand{\smallgraphcase}[2]{\begin{tabular}{c}#1\\[-2pt]\scriptsize $#2$\end{tabular}}
\setlength{\tabcolsep}{7pt}
\begin{tabular}{c*{4}{c}}
\hline
Graph & \multicolumn{4}{c}{Decoration} \\
\hline
\noalign{\vskip 0.5em}
$P_2$ & \smallgraphcase{\twograph{white}{white}}{U=\varnothing} & \smallgraphcase{\twograph{orange!65}{white}}{U=\{1\}} & & \\
$P_3$ & \smallgraphcase{\threegraph{(v1)--(v2) (v2)--(v3)}{white}{white}{white}}{U=\varnothing} & \smallgraphcase{\threegraph{(v1)--(v2) (v2)--(v3)}{orange!65}{white}{white}}{U=\{1\}} & \smallgraphcase{\threegraph{(v1)--(v2) (v2)--(v3)}{white}{orange!65}{white}}{U=\{2\}} & \smallgraphcase{\threegraph{(v1)--(v2) (v2)--(v3)}{orange!65}{white}{orange!65}}{U=\{1,3\}} \\
$K_3$ & \smallgraphcase{\threegraph{(v1)--(v2) (v2)--(v3) (v1)--(v3)}{white}{white}{white}}{U=\varnothing} & \smallgraphcase{\threegraph{(v1)--(v2) (v2)--(v3) (v1)--(v3)}{orange!65}{white}{white}}{U=\{1\}} & & \\
\hline
\end{tabular}
\caption{The inequivalent connected decorated graphs on two and three vertices that parameterize the resource states in Table~\ref{tab:robustness-exhaustive-small}. Orange vertices carry a Hadamard decoration.}
\label{tab:robustness-small-catalogue}
\end{table}

{Now, we show a strict separation between the robustness values given in Proposition \ref{cor:direct comparison of robustness}.}
{Let $D$, $S$, $C$, and $M$ denote the deterministic, stabilizer, CNC, and $p=2$ max-weight phase spaces, respectively.
 {Both $D$ and $M$ are local-Clifford orbits of reference phase-space operators;
 for $M$, the reference operator corresponds to the NBL function $\gamma^{(2)}$, defined above.}
The local phase spaces used in our analysis are
\[
L_{0}=D,\qquad
L_{1}=D\cup M,\qquad
L_{2}=S\cup D\cup M,\qquad
L_{3}=C\cup D\cup M,
\]
which satisfy the inclusions $L_{0}\subseteq L_{1}\subseteq L_{k}$ for $k=2,3$. Consequently,
\[
\mathfrak{R}_{L_{3}}(\rho)
\leq
\mathfrak{R}_{L_{2}}(\rho)
\leq
\mathfrak{R}_{L_{1}}(\rho)
\leq
\mathfrak{R}_{L_{0}}(\rho)
\]
for every quantum state $\rho$. The latter two inequalities follow directly from the corresponding inclusions of phase spaces, while the first follows from $\on{SP}_{n}\subseteq\on{CP}_{n}$.
}

{%
We compute robustness values for the MBPC resource states
\[
\rho_{G,U}
=
T^{\otimes n}H^{\otimes U}
\lvert G\rangle\!\langle G\rvert
H^{\otimes U}(T^\dagger)^{\otimes n},
\]
which are classified by inequivalent connected decorated graphs $G$, with the decorations forming independent sets $U$ that specify the qubits on which the Hadamard gate is applied. Table~\ref{tab:robustness-exhaustive-small} gives the exhaustive comparison for the resource states on two and three qubits and exhibits cases in which the $L_{3}$ robustness is strictly smaller than the CNC robustness.
{
These examples also establish strictness in
Proposition~\ref{cor:direct comparison of robustness}, where
the robustness measures are evaluated on different resource states.
Indeed, when $U=\varnothing$, we have
\[
\rho_{G,\varnothing}
=
E(G)(\ket{T}\bra{T})^{\otimes n}E(G).
\]
Since $E(G)$ is Clifford and the CNC polytope is invariant under
Clifford conjugation, we have
\[
\mathfrak{R}_C(\rho_{G,\varnothing})
=
\mathfrak{R}_C((\ket{T}\bra{T})^{\otimes n}).
\]
Thus, for $G=P_3$ and $G=K_3$, the table gives
\[
\mathfrak{R}_L(\rho_{G,\varnothing})
\leq
\mathfrak{R}_{L_3}(\rho_{G,\varnothing})
<
\mathfrak{R}_C(\rho_{G,\varnothing})
=
\mathfrak{R}_C((\ket{T}\bra{T})^{\otimes 3}).
\]
Numerically, the $L_3$ robustness values are approximately
$1.183$ and $1.190$, respectively, whereas the CNC robustness
of the product magic state is approximately $1.283$.
Consequently, the comparison between the MBPC and PBC resource
states is strict in both cases.
}

For four qubits, Table~\ref{tab:robustness-four-catalogue} displays all 24 inequivalent connected decorated graphs, while Figure~\ref{fig:robustness-four-bars} compares the robustness of the corresponding resource states in the $L_{1}$, $L_{2}$, and stabilizer phase spaces. We find that $L_{2}$ yields a strict improvement over the stabilizer phase space in 22 of the 24 cases. In the remaining two cases, the resource states are themselves stabilizer states, and hence both robustness values equal unity.%
\footnote{Interestingly, the application of $T^{\otimes4}$ does not necessarily take a decorated graph state outside the stabilizer sector. For example, resource state 5 satisfies $H^{\otimes U}\ket{G} = \ket{\mathrm{GHZ}} = ({\ket{0000}+\ket{1111}})/{\sqrt{2}}$ and applying $T^{\otimes4}$ yields $T^{\otimes4}\ket{\mathrm{GHZ}} = ({\ket{0000}-\ket{1111}})/{\sqrt{2}}$, which is again a stabilizer state.}
}
 {
For the undecorated graphs, Clifford invariance similarly gives
\[
\mathfrak{R}_S(\rho_{G,\varnothing})
=
\mathfrak{R}_S((\ket{T}\bra{T})^{\otimes 4}).
\]
Hence the strict inequalities shown for these graphs imply
\[
\mathfrak{R}_L(\rho_{G,\varnothing})
\leq
\mathfrak{R}_{L_2}(\rho_{G,\varnothing})
<
\mathfrak{R}_S((\ket{T}\bra{T})^{\otimes 4}).
\]
These four-qubit examples therefore also give a strict
comparison with stabilizer-based estimation on the corresponding
PBC resource state.
}

\renewcommand{\arraystretch}{1.35}
\begin{table}[t]
\centering
\newcommand{\fourgraph}[5]{
\begin{tikzpicture}[baseline=-0.5ex,scale=0.72]\coordinate (v1) at (-0.62,0.46);\coordinate (v3) at (0.62,0.46);\coordinate (v4) at (-0.62,-0.46);\coordinate (v2) at (0.62,-0.46);\draw[line width=0.45pt] #1;\node[circle,draw,fill=#2,minimum size=3.2mm,inner sep=0pt,font=\tiny] at (v1) {1};\node[circle,draw,fill=#3,minimum size=3.2mm,inner sep=0pt,font=\tiny] at (v2) {2};\node[circle,draw,fill=#4,minimum size=3.2mm,inner sep=0pt,font=\tiny] at (v3) {3};\node[circle,draw,fill=#5,minimum size=3.2mm,inner sep=0pt,font=\tiny] at (v4) {4};
\end{tikzpicture}}
\newcommand{\graphcase}[7]{\begin{tabular}{c}\fourgraph{#1}{#2}{#3}{#4}{#5}\\[-2pt]\scriptsize #6: $#7$\end{tabular}}
\setlength{\tabcolsep}{5pt}

\begin{tabular}{c*{5}{c}}
\hline
Graph & \multicolumn{5}{c}{State Index and Decoration} \\
\hline
\noalign{\vskip 0.5em}
$G_3$ & \graphcase{(v1)--(v4) (v2)--(v4) (v3)--(v4)}{white}{white}{white}{white}{1}{U=\varnothing}
& \graphcase{(v1)--(v4) (v2)--(v4) (v3)--(v4)}{orange!65}{white}{white}{white}{2}{U=\{1\}}
& \graphcase{(v1)--(v4) (v2)--(v4) (v3)--(v4)}{white}{white}{white}{orange!65}{3}{U=\{4\}}
& \graphcase{(v1)--(v4) (v2)--(v4) (v3)--(v4)}{orange!65}{orange!65}{white}{white}{4}{U=\{1,2\}}
& \graphcase{(v1)--(v4) (v2)--(v4) (v3)--(v4)}{orange!65}{orange!65}{orange!65}{white}{5}{U=\{1,2,3\}} \\

$P_4$ & \graphcase{(v1)--(v3) (v1)--(v4) (v2)--(v4)}{white}{white}{white}{white}{6}{U=\varnothing}
& \graphcase{(v1)--(v3) (v1)--(v4) (v2)--(v4)}{orange!65}{white}{white}{white}{7}{U=\{1\}}
& \graphcase{(v1)--(v3) (v1)--(v4) (v2)--(v4)}{white}{orange!65}{white}{white}{8}{U=\{2\}}
& \graphcase{(v1)--(v3) (v1)--(v4) (v2)--(v4)}{orange!65}{orange!65}{white}{white}{9}{U=\{1,2\}}
& \graphcase{(v1)--(v3) (v1)--(v4) (v2)--(v4)}{white}{orange!65}{orange!65}{white}{10}{U=\{2,3\}} \\

$G_2$
& \graphcase{(v1)--(v3) (v1)--(v4) (v2)--(v4) (v3)--(v4)}{white}{white}{white}{white}{11}{U=\varnothing}
& \graphcase{(v1)--(v3) (v1)--(v4) (v2)--(v4) (v3)--(v4)}{orange!65}{white}{white}{white}{12}{U=\{1\}}
& \graphcase{(v1)--(v3) (v1)--(v4) (v2)--(v4) (v3)--(v4)}{white}{orange!65}{white}{white}{13}{U=\{2\}}
& \graphcase{(v1)--(v3) (v1)--(v4) (v2)--(v4) (v3)--(v4)}{white}{white}{white}{orange!65}{14}{U=\{4\}}
& \graphcase{(v1)--(v3) (v1)--(v4) (v2)--(v4) (v3)--(v4)}{orange!65}{orange!65}{white}{white}{15}{U=\{1,2\}} \\

$C_4$
& \graphcase{(v1)--(v3) (v1)--(v4) (v2)--(v3) (v2)--(v4)}{white}{white}{white}{white}{16}{U=\varnothing}
& \graphcase{(v1)--(v3) (v1)--(v4) (v2)--(v3) (v2)--(v4)}{orange!65}{white}{white}{white}{17}{U=\{1\}}
& \graphcase{(v1)--(v3) (v1)--(v4) (v2)--(v3) (v2)--(v4)}{orange!65}{orange!65}{white}{white}{18}{U=\{1,2\}} & & \\

$G_1$
& \graphcase{(v1)--(v3) (v1)--(v4) (v2)--(v3) (v2)--(v4) (v3)--(v4)}{white}{white}{white}{white}{19}{U=\varnothing}
& \graphcase{(v1)--(v3) (v1)--(v4) (v2)--(v3) (v2)--(v4) (v3)--(v4)}{orange!65}{white}{white}{white}{20}{U=\{1\}}
& \graphcase{(v1)--(v3) (v1)--(v4) (v2)--(v3) (v2)--(v4) (v3)--(v4)}{white}{white}{orange!65}{white}{21}{U=\{3\}}
& \graphcase{(v1)--(v3) (v1)--(v4) (v2)--(v3) (v2)--(v4) (v3)--(v4)}{orange!65}{orange!65}{white}{white}{22}{U=\{1,2\}} & \\

$K_4$
& \graphcase{(v1)--(v2) (v1)--(v3) (v1)--(v4) (v2)--(v3) (v2)--(v4) (v3)--(v4)}{white}{white}{white}{white}{23}{U=\varnothing}
& \graphcase{(v1)--(v2) (v1)--(v3) (v1)--(v4) (v2)--(v3) (v2)--(v4) (v3)--(v4)}{orange!65}{white}{white}{white}{24}{U=\{1\}} & & & \\
\hline
\end{tabular}
\caption{The 24 inequivalent connected decorated four-vertex graphs. Orange vertices carry a Hadamard decoration. The number preceding each decoration is the state index on the horizontal axis of the Figure~\ref{fig:robustness-four-bars}.
}
\label{tab:robustness-four-catalogue}
\end{table}

\begin{figure}[t]
\centering
\begin{tikzpicture}
\begin{axis}[ybar=0pt,bar width=4pt,width=\textwidth,height=7.2cm,xmin=0.5,xmax=24.5,ymin=0,ymax=3.1,xtick={1,...,24},xlabel={State Index},ylabel={Robustness},ymajorgrids=true,legend columns=3,legend style={at={(0.5,1.02)},anchor=south,draw=none},enlarge x limits=false]
\addplot+[fill=blue!55,draw=blue] coordinates {(1,1.445432) (2,1.457107) (3,1.195652) (4,1.353553) (5,1.500000) (6,1.957107) (7,1.804738) (8,1.707107) (9,1.810660) (10,2.000000) (11,1.935660) (12,1.804738) (13,2.164214) (14,1.727273) (15,1.810660) (16,1.560660) (17,1.804738) (18,1.810660) (19,1.737437) (20,1.804738) (21,1.727273) (22,1.810660) (23,1.260200) (24,1.195652)};
\addlegendentry{$\mathfrak R_{L_1}$}
\addplot+[fill=red!55,draw=red] coordinates {(1,1.445432) (2,1.414214) (3,1.195652) (4,1.353553) (5,1.000000) (6,1.957107) (7,1.804738) (8,1.707107) (9,1.471405) (10,1.000000) (11,1.913583) (12,1.804738) (13,1.530330) (14,1.727273) (15,1.471405) (16,1.560660) (17,1.804738) (18,1.471405) (19,1.714855) (20,1.804738) (21,1.727273) (22,1.471405) (23,1.260200) (24,1.195652)};
\addlegendentry{$\mathfrak R_{L_2}$}
\addplot+[fill=orange!65,draw=brown] coordinates {(1,2.862742) (2,1.747547) (3,2.555556) (4,1.747547) (5,1.000000) (6,2.862742) (7,2.800610) (8,1.747547) (9,1.747547) (10,1.000000) (11,2.862742) (12,2.800610) (13,1.747547) (14,2.555556) (15,1.747547) (16,2.862742) (17,2.800610) (18,1.747547) (19,2.862742) (20,2.800610) (21,2.555556) (22,1.747547) (23,2.862742) (24,2.555556)};
\addlegendentry{$\mathfrak R_S$}
\end{axis}
\end{tikzpicture}
\caption{Robustness values for all 24 inequivalent connected four-qubit resource states with independent Hadamard decorations depicted in Table~\ref{tab:robustness-four-catalogue}. The phase spaces satisfy $L_1\subseteq L_2$ as well as $S\subseteq L_2$, so that $\mathfrak R_{L_2}\leq\mathfrak R_{L_1}$ and $\mathfrak R_{L_2}\leq\mathfrak R_S$ in every case. {The latter inequality is strict for 22 states; states 5 and 10 satisfy $\mathfrak R_{L_2}=\mathfrak R_S=1$.}}
\label{fig:robustness-four-bars}
\end{figure}

\section{{Conclusion}}
\label{sec:conclusion}

{In this work, we introduce a new classical simulator for
quantum computation that uses the geometry of operator polytopes.
We first formalize the notion of polyhedral classical simulation
and then construct the local Pauli simulator for measurement-based
Pauli computation (MBPC), a form of measurement-based quantum
computation. Polyhedral simulators can represent quantum states
as probability distributions over the vertices of a polytope,
and they update these distributions under quantum operations
by sampling from these vertices. Thus, polyhedral simulators
translate the task of classical simulation into problems of
polyhedral geometry and sampling, connecting quantum computational
complexity with polyhedral complexity. The first example of a polyhedral simulator, the Pauli ($\Lambda$)
simulator, was introduced in~\cite{zurel2020hidden} for
Pauli-based computation (PBC), a particular form of
quantum computation with magic states. {Our local Pauli simulator operates on a larger polytope than
the Pauli simulator. For explicit resource states, we obtain
strictly smaller sample-complexity bounds than CNC-based
simulation.}}

{The complexity of polyhedral simulators consists of two parts:
(1) sample complexity and (2) per-sample complexity. The sample
complexity is determined by the robustness measure of the initial quantum
state with respect to the polytope. The per-sample complexity is determined
by the cost of sampling from the initial distribution and the updated vertex operators after
each quantum operation.
Constructing these distributions requires finding convex
decompositions of operators inside the polytope in terms of
its vertices. In the direct approach, one first enumerates
the vertices and then solves the corresponding convex
decomposition problems. Vertex enumeration is therefore
a central computational obstacle in this approach, in addition
to the cost of constructing and sampling the decompositions.}
No known algorithm efficiently solves the vertex enumeration problem for all polytope families \cite{avis1995good}.
Efficiency here means a solution that is polynomial in
the number of facets, vertices and the dimension of the polytope.
We expect vertex enumeration for {the local Pauli polytope} {$\LP_{n}$} to be \textit{more efficient} in \( n \) compared to {the Pauli polytope} \( \Lambda_{n} \).
{B}oth
polytopes have the same
dimension, scaling as \( 4^n - 1 \), {but} the number of facets
scale as  $2^{O(n)}$ and $2^{O(n^2)}$, respectively.
While this doesn't guarantee that the number of vertices will be less in the local case {\cite{avis1995good}}, there is evidence supporting this. For example, for \( n = 2 \), we have \( |S_{2}^{\loc}| = 36 \) and \( |S_2| = 60 \), with the number of vertices \(1,408 \) and \(22,320 \), respectively. {Initial numerical results suggest that this continues to hold for $n=3$.}

{We identify the local Pauli polytope with the non-signaling
polytope of the $n$-partite Bell scenario with three binary-outcome
measurements per party, a fundamental polytope in the study of
Bell inequalities in quantum foundations. The specialization
of our polyhedral simulator to the Bell polytope, which we call
the deterministic vertex simulator, has efficient update rules.
On the other hand, the Pauli simulator has a well-known efficiently simulatable sector conssiting of the closed non-contextual (CNC) operators.
The deterministic polytope and the CNC polytope, whose operators
are simulated by the phase-space tableau algorithm
of~\cite{ipek2025phase}, provide overlapping regions for classical
simulation, neither of which contains the other in general.
Combining deterministic vertices with CNC operators therefore
enlarges the simulation state space beyond either construction
alone, while retaining efficient update rules.}

{We further extend the simulation state space by introducing
locally closed operators as a generalization of CNC operators.
Besides deterministic vertices, these include max-weight vertices
specified by efficiently computable non-bipartite-linear functions,
providing new families of efficiently representable vertices. Combining deterministic vertices, CNC operators, and these
max-weight vertices enlarges the available efficiently simulatable sector.
Efficient exact sampling additionally requires an efficiently
samplable convex decomposition of the resource state. For
quasi-probability estimation, both the cost of sampling the
initial decomposition and its $1$-norm must be controlled. Our reduced resource construction converts a circuit with
$t$ $T$ gates into an MBPC on $t$ resource qubits, enabling a
direct comparison with PBC on the same number of qubits.
Proposition~\ref{cor:direct comparison of robustness} and the
three-qubit examples show that the local robustness of the
reduced MBPC resource can be strictly smaller than the CNC
robustness of the corresponding product magic resource in PBC. These strict robustness separations therefore yield smaller
sample complexity bounds for the local Pauli simulator.}

 From a foundational perspective, contextuality is also a resource
in our model since any state in the Bell polytope
{with an efficiently samplable convex decomposition over
deterministic vertices} is efficiently simulatable.
As in earlier work~\cite{raussendorf2017contextuality},
{this connects non-contextuality with efficient classical
simulation, subject to the initialization requirement}.
However, in our model there are contextual states that can
be simulated efficiently.
On the other hand, the power of correlations in the MBQC model maps nicely into our model when the measurements are local Pauli {measurements}. This intersection provides a new playground to see how computational advantage relates to various computational models, including QCM and MBQC, and the corresponding monotones, robustness, and contextual fraction \cite{raussendorf2013contextuality,abramsky2017contextual}.
{The recent topological framework for contextuality, based on the theory of simplicial distributions \cite{okay2022simplicial}, captures the operator-theoretic polytopes used in simulation, including the Pauli and local Pauli polytopes. This framework provides a natural setting for further exploration and analysis.}

Our results open up new avenues for analyzing and simulating quantum computation using tools from polyhedral geometry, quantum foundations, and generalized probabilistic theories. The techniques developed here lay a foundation for further investigations into the role of contextuality, entanglement, and magic in defining the classical simulability frontier in quantum computation.

\paragraph{Acknowledgments}
This work is supported by the Digital Horizon Europe project FoQaCiA, GA no.
101070558, and the US Air Force Office of Scientific Research
under award number FA9550-21-1-0002 {and FA9550-24-1-0257}.
 We thank Robert Raussendorf for discussions.
 {{Most of the second author's contribution to this work was carried out
while he was an undergraduate student in Computer Engineering at
Bilkent University.} Most of the third author's contribution to this work was carried out during his postdoctoral appointment at Bilkent University.}

\paragraph{Conflict of interest}
The authors declare no competing interests.

\paragraph{Data and code availability}
The numerical data and code supporting the findings of this study
are available in the LocalPauli repository~\cite{LocalPauli}.

\paragraph{Ethics statement}
This study does not involve human participants or animal subjects. Ethical approval was not required.


\bibliography{bib}
\bibliographystyle{ieeetr}

\end{document}